\documentclass[aps,prx,reprint,superscriptaddress, twocolumns]{revtex4-2}
\usepackage{amsmath,amssymb}
\usepackage{bbold}
\usepackage{float}
\usepackage{graphicx}
\usepackage{dcolumn}
\usepackage{bm}
\usepackage[colorlinks]{hyperref}
\usepackage{braket}
\usepackage{mathtools}
\usepackage{comment}
\usepackage{xcolor}
\usepackage{soul}
\usepackage{color}
\usepackage{physics}
\usepackage{enumerate}
\usepackage{subcaption}
\usepackage{tikz-cd} 
\usepackage{cleveref}
\crefmultiformat{equation}{(#2#1#3)}%
{ and~(#2#1#3)}{, (#2#1#3)}{ and~(#2#1#3)}

\colorlet{alena}{blue!80!gray}
\colorlet{roman}{green!40!gray}

\newtheorem{thm}{Theorem}
\newtheorem{dfn}[thm]{Definition}

\newcommand{\R}{{\rm I\!R}}
\newcommand{\TQOptimaX}{our implementation of SAMO-COBRA}

\begin{document}

\title{Multi-objective optimization and quantum hybridization \\ of equivariant deep learning interatomic potentials}

\author{G.~Laskaris}
\affiliation{Terra Quantum AG, Kornhausstrasse 25, 9000 St. Gallen, Switzerland}
\affiliation{LIACS, Leiden University, Netherlands}

\author{D.~Morozov}%
\affiliation{Terra Quantum AG, Kornhausstrasse 25, 9000 St. Gallen, Switzerland}
\affiliation{Nanoscience Center and Department of Chemistry, University of Jyväskylä, P.O. Box 35, 40014 Jyväskylä, Finland}

\author{D.~Tarpanov}%
\affiliation{Terra Quantum AG, Kornhausstrasse 25, 9000 St. Gallen, Switzerland}

\author{A.~Seth}%
\affiliation{Department of Materials Engineering, Indian Institute of Science, Bengaluru 560012, India}

\author{J.~Procelewska}%
\affiliation{Schaeffler Technologies AG \& Co. KG, Herzogenaurach, Germany}

\author{G.~Sai~Gautam}%
\affiliation{Department of Materials Engineering, Indian Institute of Science, Bengaluru 560012, India}

\author{A.~Sagingalieva}%
\affiliation{Terra Quantum AG, Kornhausstrasse 25, 9000 St. Gallen, Switzerland}

\author{R.~Brasher}%
\affiliation{Terra Quantum AG, Kornhausstrasse 25, 9000 St. Gallen, Switzerland}

\author{A.~Melnikov}%
\affiliation{Terra Quantum AG, Kornhausstrasse 25, 9000 St. Gallen, Switzerland}

\begin{abstract}
 Allegro is a machine learning interatomic potential model designed to predict atomic properties in molecules using E(3) equivariant neural networks. When training this model, there tends to be a trade-off between accuracy and inference time. For this reason, we apply multi-objective hyperparameter optimization to both objectives. Additionally, we experiment with modified architectures by constructing variants of Allegro: one extended with additional classical layers and one incorporating quantum-classical hybrid layers. We evaluate all models on QM9, rMD17-aspirin, rMD17-benzene, and a self-generated dataset of copper-lithium structures. As results, both variants surpass Allegro in force prediction accuracy across multiple datasets. The classical variant consistently improves over the baseline, while the quantum-classical hybrid variant achieves the best overall force prediction accuracy on the Cu-Li dataset, where it was fully optimized, outperforming the classical variant by approximately 13\%. Notably, the hybrid variant also achieves competitive results on the remaining datasets despite using hyperparameters transferred from Cu-Li without dataset-specific optimization, suggesting that quantum-classical hybridization is a promising direction for enhancing MLIP architectures.
\end{abstract}

\maketitle

\begin{figure}[b!]
\noindent\fbox{\parbox{\dimexpr\columnwidth-2\fboxsep-2\fboxrule\relax}{\raggedright Please check the published version, which includes all the latest additions and corrections: Comput. Mater. Sci. 270:114742, 2026, DOI: \href{https://doi.org/10.1016/j.commatsci.2026.114742}{10.1016/j.commatsci.2026.114742}}}
\end{figure}

\section{\label{sec:introduction} Introduction}

Machine learning interatomic potential (MLIP) models \cite{mlip_1,mlip_2,mlip_3} are machine learning models trained to predict energies, stresses, forces of molecular systems. They map the atomic structure of molecules into their potential energies. An MLIP model, in comparison with density functional theory (DFT) \cite{dft_1,dft_2}, reduces computational cost and lowers the complexity of energy calculations, while approaching DFT accuracy. Some widely used MLIP models are SchNet \cite{schnet_1}, Gaussian Approximation Potentials (GAP) \cite{gap}, Smooth Overlap of Atomic Positions - GAP (SOAP-GAP) \cite{soap-gap}, Atomic Cluster Expansion (ACE) \cite{ACE, PACE} and NequiP model \cite{nequip_paper}.

Here we conduct a methodical multi-objective optimization search of hyperparameters for an improved version of the aforementioned NequiP model which is the Allegro model \cite{allegro_paper}. Allegro yields significant improvements over NequiP \cite{allegro_paper}, especially on large-scale systems. It allows for parallel computations by dropping the message passing propagation of the information and substituting it with a series of tensor products of learned equivariant representations. We optimize both the objectives of the accuracy and the inference time. In addition, we perform a network architecture search for variants of the Allegro model, including quantum-classical hybrid models. We extend our hyperparameter optimization search on additional Allegro model variants, such as an extension of Allegro with additional multi-layer perceptron (MLP) layers along with quantum depth infused (QDI) layers \cite{sagingalieva2023hybrid, sagingalieva2025photovoltaic, lusnig2024hybrid, sedykh2024hybrid, anoshin2024hybrid, lee2025steel}.

There have been other areas where researchers have found success by inserting quantum layers in purely classical models, a process called, quantum-classical hybridization \cite{sagingalieva2025hybrid, sagingalieva2023hybrid, tsurkan2026hybrid, kurkin2025forecasting, lopatkin2026selector}.

To ensure that we find the best fitting hyperparameters for all variant Allegro models, we use a state-of-the-art multi-objective optimization algorithm, \TQOptimaX{} \cite{samo-cobra} for the hyperparameter optimization. This algorithm runs multiple evaluations of our objective function and at the same time it searches and optimizes $6-9$ hyperparameters depending on the type of Allegro model. We evaluate the performance of variant Allegro models on a plethora of datasets. One of them is an unpublished and self-generated dataset, and three widely used public datasets. The unpublished dataset includes inorganic compounds and specifically copper and lithium atoms. The public datasets are: $(1)$ QM9 \cite{qm9_1}, which has $133885$ different molecules, produced by all the stable combinations of up to $9$ atoms of C, N, O and F, and $(2)$ the revised-MD17 (rMD17) dataset, from which we use the Aspirin and Benzene subsets \cite{rmd17}. We train the models to predict molecular energies from QM9 and both energies and forces for Cu-Li, Aspirin and Benzene datasets.

\section{\label{sec:preliminaries} Preliminaries}

In this section, we outline key concepts of our work. We discuss the importance of equivariance and symmetries in graph neural networks (GNNs), introduce \TQOptimaX, and describe the QDI layer enhancements of the Allegro model.

\subsection{\label{sec:allegro} Allegro model}

In the context of molecular dynamics and the prediction of energies and forces, selecting an appropriate symmetry group $G$ is vital. Equivariance arises naturally by the geometrical properties and symmetries of molecular systems \cite{equivariance_in_molecules}, and thus comprises a mechanism that boosts the performance of MLIP \cite{nequip_paper, allegro_paper}.

The benefit of equivariance derives from the ability to reduce the space of possible weight matrices for a multilayer perceptron, in contrast with other MLIP models which require a large amount of configurations to achieve high-fidelity results \cite{mlip_with_large_data_1, mlip_with_large_data_2}.

Allegro model \cite{allegro_paper,allegro_application,allegro_application_2} is a Graph Neural Network (GNN) that utilizes internal symmetries to build an equivariance neural network \cite{butterfly_paper, equivariant_cnn, e(n)_equivariant_gnns, equivariance_weights_relation} with a reduced number of trainable parameters.

The demand for more accurate and sophisticated neural networks leads us to exploit different characteristics and properties. The notion of equivariance of a function to a group is characterized as a generalization of the invariance \cite{butterfly_paper}. Regarding Allegro, the group over which the network is equivariant is E$(3)$. The group operations of E$(3)$ encompass rotations, translations in $3$ dimensions, and reflections.

\begin{dfn}[Equivariance]
    Suppose the sets $\mathcal{X}, \mathcal{Y}$ and a function $f:\mathcal{X} \rightarrow \mathcal{Y}$. Then for a group $G$ and a transformation $g \in G$ , if:

    \begin{equation}
        f(T_g x) = T^\prime_g f(x), \quad \forall \: x\in \mathcal{X}, g\in G \label{eq:equivariance}
    \end{equation}
    
    Then $f$ is equivariant to group $G$.
\end{dfn}

 Note that $T_g$ and $T_g^\prime$ are the representation operators of $g$ in $\mathcal{X}$ and $\mathcal{Y}$ spaces, respectively. Also, these operators do not have to be equal but $T$ has to be a linear representation of G \cite{equivariant_cnn}. In case of $T^\prime_g$ being the identity transformation then we fall to the special case of $f$ being invariant to group $G$.

For our case, the commutative diagram below, represents graphically the equivariance.
 
\[ \begin{tikzcd}[column sep=7em, row sep =4.8em]
\mathcal{X} \arrow{r}{T_g} \arrow[swap]{d}{f} & \mathcal{X} \arrow{d}{f} \\%
\mathcal{Y} \arrow{r}{T^\prime_g}& \mathcal{Y}
\end{tikzcd}
\]

\begin{dfn}[$k$-layered multilayer perceptron \cite{butterfly_paper}]
    Let $\mathcal{A} \in \R^{n_{i-1} \times n_i}, \: i=\{1,\ldots,k\}$ be all possible weights matrices, $\mathbf{b}_i 
    \in \R^{n_i}, \: i=\{1,\ldots,k-1\}$ be the bias terms (assuming no bias on the output layer) and $\sigma:\R \rightarrow \R$ be a continuous activation function. Let $G$ be a group with representation $\rho_i:G\rightarrow GL(n_i), \: i=\{0,\ldots,k\}$. Then the $k$-layered multilayer perceptron $f:\R^{n_0}\rightarrow \R^{n_k}$ is

    \[
    f(x) = \mathcal{A}_k\sigma_{\mathbf{b}_{k-1}}\mathcal{A}_{k-1}\sigma_{\mathbf{b}_{k-2}}\ldots\sigma_{\mathbf{b}_1}\mathcal{A}_1 x
    \]

    Then $f$ is equivariant to G with respect to $\rho_i$ if 
    \begin{align}
        \mathcal{A}_i \rho_{i-1}(g) = \rho_i(g) \mathcal{A}_i, \quad \text{and}\label{eq:def1.1}\\
        \sigma_{\mathbf{b}_i} \rho_i(g) = \rho_i(g) \sigma_{\mathbf{b}_i}, \quad \forall g \in G \label{eq:def1.2}
    \end{align}

    With $\sigma_{\mathbf{b}}:\R^{n_i}\rightarrow\R^{n_i}, \:  \sigma_{\mathbf{b}}(x) = \sigma(x+\mathbf{b})$ being a point-wise nonlinearity.
\end{dfn}

Equations~\cref{eq:def1.1,eq:def1.2} are both derived from equivariance definition replacing the representation operator with the representation $\rho_i(g)$ of $i$-th layer. Then by following the proof from \cite{butterfly_paper} we can show that $f(\rho_0(g)x) = \rho_k(g)f(x)$. Equation~\eqref{eq:def1.2} is always satisfied by any point-wise nonlinearity while all components of $b_i$ are equal to the same real number. While this is true, it is evident how equivariance limits the possible choices for weights matrices, and thus reduces the complexity of our problem. As a consequence, neural networks that utilize the equivariance symmetry are capable on achieving high-fidelity results with significant less data \cite{nequip_paper,allegro_paper}.

\subsection{\label{sec:cva} Details of \TQOptimaX}

We use \TQOptimaX{} as a multi-objective optimization algorithm that attempts to find the set of hyperparameters that minimize all the (expensive) objective functions we want to optimize. Particularly, \TQOptimaX{} succeeds on maximizing the hyper-volume in the solution space by minimizing the evaluations of the objective functions. Formally,

\begin{equation}
    \begin{aligned}
       & \text{minimize} \: \mathbf{f(x)} = \Omega \rightarrow \mathbb{R}^k \: , \: \mathbf{f(x)} = \{f_1(\mathbf{x}),\ldots , f_k(\mathbf{x})\} \\
       & \text{subject to} \: g_i(\mathbf{x}) \leq 0 \: , \: \forall i \in \{1,\ldots,m\} , \mathbf{x} \in \Omega \subset \mathbb{R}^d
    \end{aligned}
\end{equation}

where $k$ is the number of objective functions we want to minimize, $m$ is the number of constraints and $d$ the dimensionality of the problem. In our case, we want to minimize two objective functions, namely mean absolute error (MAE) of forces with respect to the validation set and the inference time of the entire test set.

We use \TQOptimaX{} to maximize the hypervolume ($Hv$) \cite{samo-cobra} of the region in the objective space dominated by a given set of solutions.

\subsection{\label{sec:qdi} Quantum depth-infused layer}
\begin{figure}[ht]
    \includegraphics[width=1\linewidth]{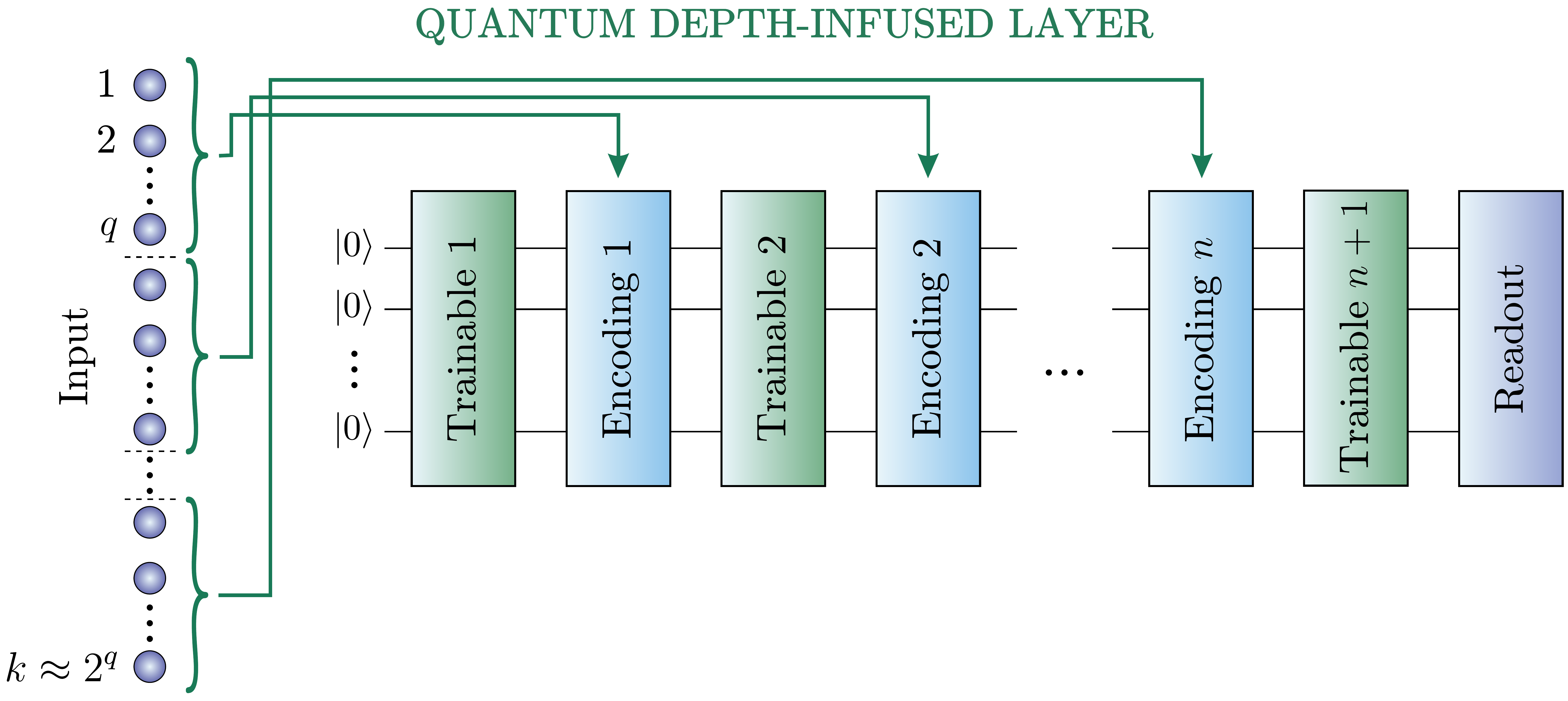}
    \caption{Quantum depth-infused layer}
    \label{fig:qdi}
\end{figure}

In the proposed ML model, a quantum layer, implemented as a specialized variational quantum circuit (VQC) known as the QDI layer~\cite{sagingalieva2023hybrid}, is leveraged to manage a large number of input features with a limited qubit count. The structure of this quantum layer is illustrated in Fig.~\ref{fig:qdi}. Unlike traditional approaches where each feature is assigned its own qubit, the QDI layer utilizes a data re-uploading strategy. This method sequentially encodes features onto a lattice structure. By incorporating entangling gates and variational quantum layers across encoding sections, the model can capture complex, non-linear patterns and correlations. This design enhances the model’s representational power.

\begin{figure*}[ht]
    \centering
    \begin{subfigure}[b]{0.45\linewidth}
        \centering
        \includegraphics[width=\linewidth]{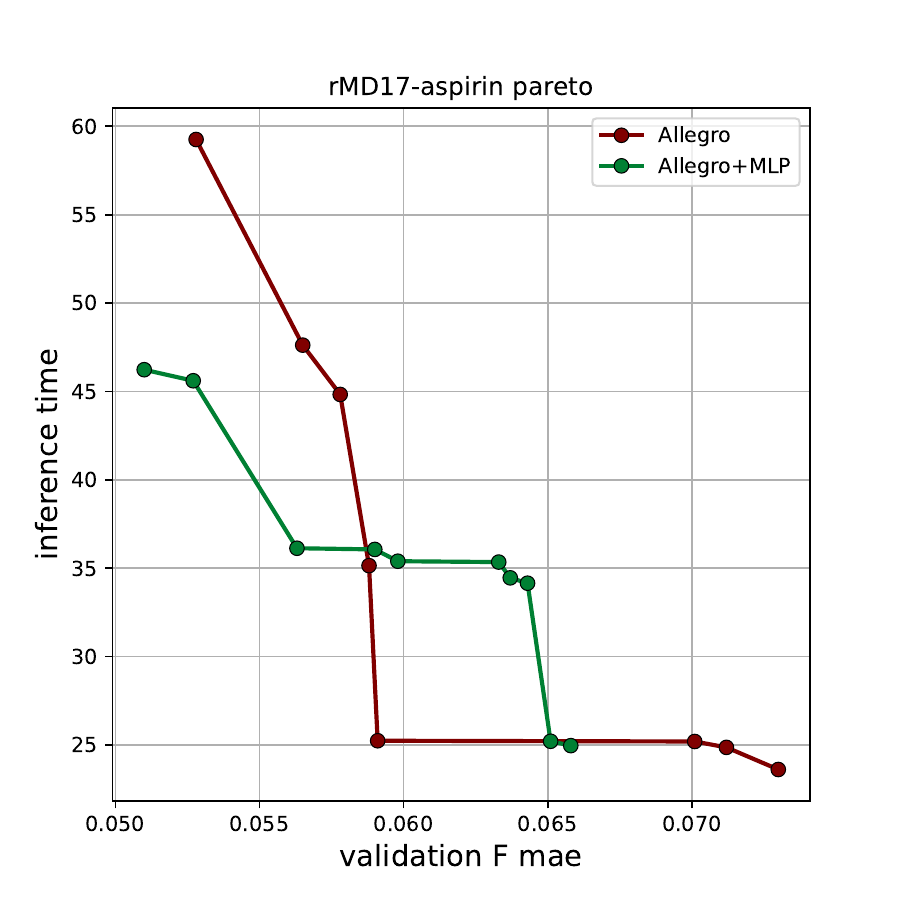} \label{subfig:pareto_aspirin}
        \caption{Pareto front of the rMD17-aspirin dataset}
    \end{subfigure}
    \begin{subfigure}[b]{0.45\linewidth}
        \centering
        \includegraphics[width=\linewidth]{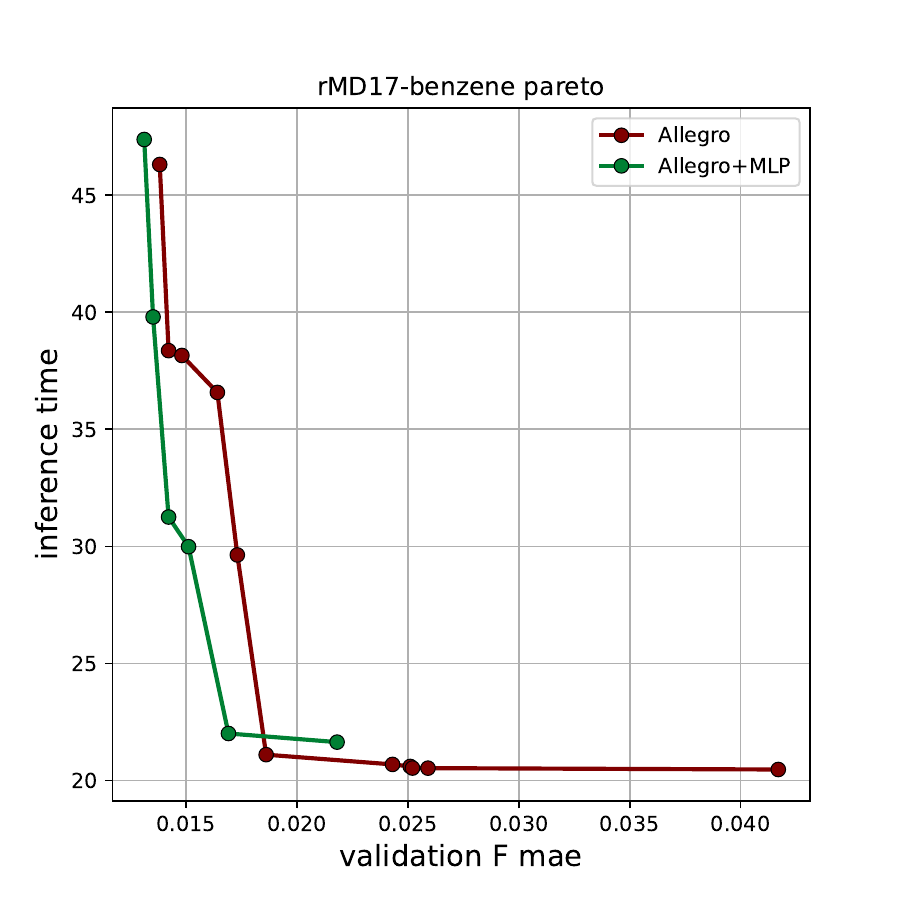} \label{subfig:pareto_benzene}
        \caption{Pareto front of the rMD17-benzene dataset}
    \end{subfigure}
    
    \vspace{1em} 
    
    \begin{subfigure}[b]{0.45\linewidth}
        \centering
        \includegraphics[width=\linewidth]{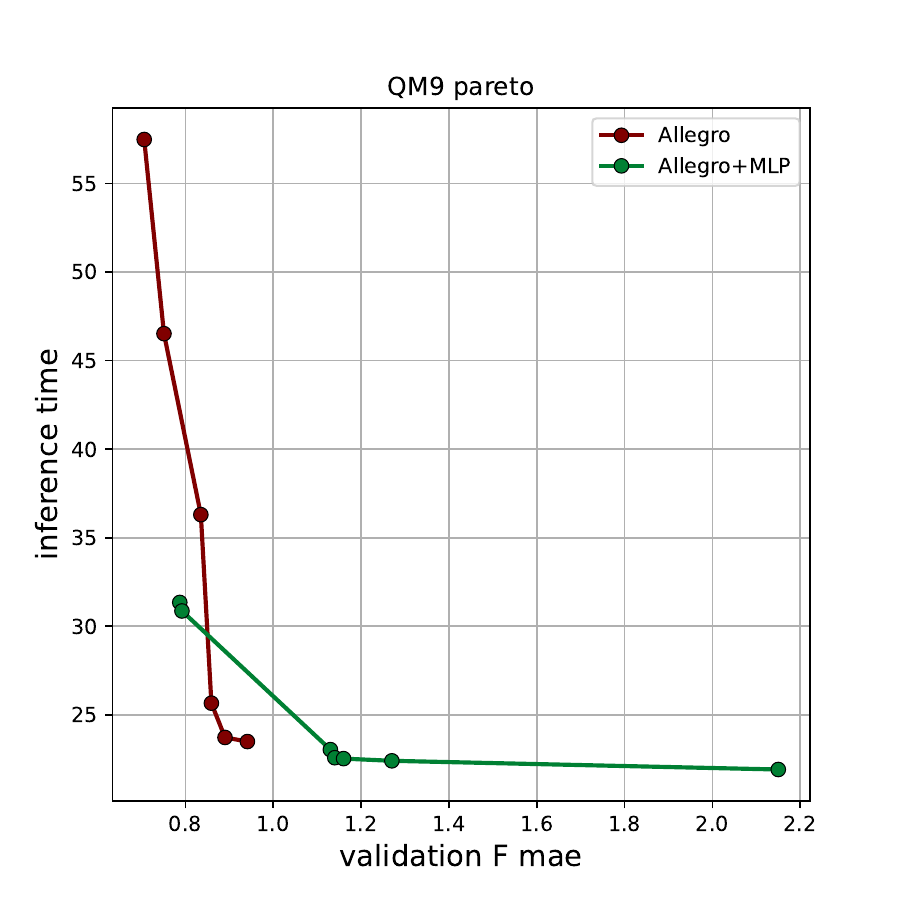} \label{subfig:pareto_qm9}
        \caption{Pareto front of the QM9 dataset}
    \end{subfigure}
    \begin{subfigure}[b]{0.41\linewidth}
        \centering
        \includegraphics[width=\linewidth]{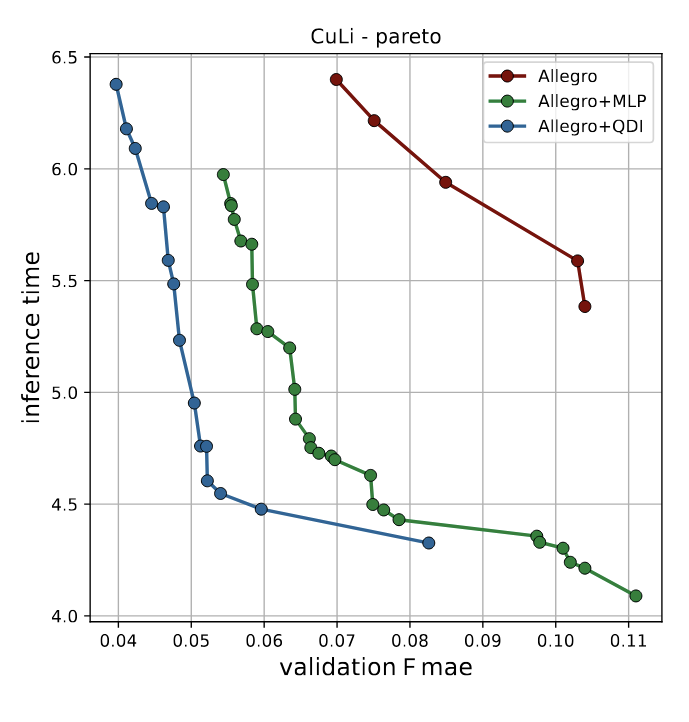} \label{subfig:pareto_culi}
        \caption{Pareto front of the Cu-Li dataset}
    \end{subfigure}
    
    \caption{Pareto fronts as they generated by \TQOptimaX{} during multi-objective optimization of the Allegro model and its variants, minimizing both the forces mean absolute errors on the validation set in (eV/\AA) and the inference time for each dataset tested. The x-axis shows the force mean absolute error (F MAE) on the validation set in eV/\AA{}, and the y-axis shows the total inference time on the entire test set in seconds.}
    \label{fig:pareto}
\end{figure*}

\begin{table*}[htbp]
    \centering
    \begin{tabular}{c|ccccccc}
         Allegro        & $r_{max}$ & lr & batch & P & Layers & edge MLP \\ \hline
         rMD17 aspirin  & 7.3 & 5.36e-03 & 3 & 6.9 & 4 & 32 \\
         rMD17 benzene  & 12.0 & 1e-03 & 1 & 5.0 & 3 & 128 \\
         QM9            & 12.0 & 1e-03 & 5 & 10.0 & 3 & 32 \\
         CuLi           & 12.0 & 2e-03 & 2 & 5.0 & 2 & 128 \\
    \end{tabular}
    \caption{Best parameters from Allegro HPO}
    \label{tab:allegro_hpo_params}
\end{table*}

\begin{table*}[htbp]
    \centering
    \begin{tabular}{c|ccccccc}
         Allegro+MLP    & $r_{max}$ & lr & batch & P & Layers & edge MLP & final MLP  \\ \hline
         rMD17 aspirin  & 7.4 & 2.93e-03 & 2 & 6.7 & 3 & 16 & [1024,1024] \\
         rMD17 benzene  & 12.0 & 1e-03 & 1 & 5.0 & 3 & 128 & [32,4] \\
         QM9            & 6.2 & 1e-03 & 2 & 5.0 & 2 & 64 & [32,16] \\
         CuLi           & 5.0 & 9.35e-04 & 1 & 10.3 & 3 & 16 & [4,16] \\
    \end{tabular}
    \caption{Best parameters from Allegro+MLP HPO}
    \label{tab:allegro_mlp_hpo_params}
\end{table*}

\begin{table*}[htbp]
    \centering
    \begin{tabular}{c|ccccccccc}
         Allegro+QDI    & $r_\mathrm{max}$ & lr    & batch & P & Layers & edge MLP & final MLP & QDI depth & QDI input features \\ \hline
         rMD17 aspirin  & 9.5       & 1e-03 & 1     & 6.0 & 4      & 16       & 4               & 4         & 21 \\
         rMD17 benzene  & 12.0 & 1e-03 & 1 & 5.0 & 3 & 32 & 16 & 1 & 4 \\
         QM9            & 12.0 & 1e-03 & 5 & 10.0 & 3 & 32 & 128 & 3 & 4 \\
         CuLi & 5.0 & 4.02e-03 & 5 & 12.0 & 3 & 4 & 128 & 3 & 4 \\
    \end{tabular}
    \caption{Best parameters from Allegro+QDI HPO}
    \label{tab:allegro_qdi_hpo_params}
\end{table*}

\section{\label{sec:Implementation} Implementation}

For the implementation and the experiments of this work, we generated our own Cu-Li dataset from DFT calculations, we preprocessed all datasets before running any training, built all the necessary variant Allegro models, namely Allegro$+$MLP and Allegro$+$QDI and used a single T4 GPU with 16GB of memory to run the multi-objective optimization using \TQOptimaX{} as our optimization algorithm.   

\subsection{\label{sec:data} Data preprocessing}

The choice of the datasets serves multiple purposes. Training the models on QM9 allows us to explore the potential of these predictive models on a diverse set of organic molecules. We pre-processed QM9 by excluding the $3054$ molecules which failed the geometry consistency check from the total $133885$ molecules \cite{qm9_1}. Additionally, from all molecules' features, we isolated the energy values used for the training.

For more focused experiments we used rMD17 and specifically the aspirin and benzene subsets. For consistency with prior work, we converted all the energies from $kcal/mol$ to $eV$/atom, since we compare our results with the literature where energies are usually reported in $eV$ or $meV$.

Finally, to evaluate Allegro model and its variants on inorganic compounds, we generated a dataset consisting of copper and lithium atoms. This dataset consists of multiple $XYZ$-formatted slabs which form a surface of copper atoms and their respective energy.

\subsection{\label{sec:data-gen} Dataset generation}

The DFT-based training dataset was constructed using a diverse set of $11,635$ structures, enabling the trained MLIP to accurately model complex materials behavior in different systems. We first generated an initial set of 13 symmetrically distinct slabs using \textit{pymatgen} \cite{pyamtgen} for Li and Cu each up to the \{333\} Miller indices. The thickness of the Li and Cu slabs was fixed as $20$~Å and $30$~Å by convergence of their surface energies respectively with a vacuum spacing of 15~\AA{} on both sides to prevent periodic boundary interactions. To distinguish the bulk behavior of the slab from its surface, we only allow the atoms in the first and last two layers of each slab to relax during the DFT calculations, while immobilizing the rest of the atoms. These relaxed configurations were then used to create a wide variety of slab and interface-based structures. 

Firstly, we generate structures with surface vacancies and adatoms to capture the effect of surface modifications on these slabs. 34 structures with surface vacancies were created by removing the furthest possible surface atoms having the same coordination number in each slab. To capture defect concentration, we further generated surface vacancies in $2\times 2\times 1$ supercells of the slabs leading to an additional $34$ structures. $537$ structures with adatoms were generated by placing Li and Cu adatoms at distinct sites (hollow, top and bottom) at $2$ Å on top of the $2\times 2\times 1$ supercells of each slab and allowing the structures to relax in order to capture the adsorption behavior on the surfaces. 

Lastly, we generated a set of interfaces and melt-quench based amorphous interfacial structures to enable the potential to accurately model high-temperature environments with complex interactions between Li and Cu atoms. We used \textit{pymatgen} to create coherent interfaces between the Li and Cu slabs following the approach suggested by Zur and Mcgill \cite{Zur_Mcgill}. We started by initially transforming Li and Cu slabs into their respective supercells such that their surface areas are similar within a defined tolerance of $9$\%. For each pair of Li and Cu slabs, we found the smallest possible combination of such supercells and positioned them with a gap of $3$, thus creating geometrically viable interfaces. Due to the enormous number of interfaces that can be generated via this process, we constrained our dataset to having interfaces generated by the Li and Cu slabs sharing the same Miller indices. A melt-quench approach via ab initio molecular dynamics (AIMD) simulations was subsequently used to create amorphized interfacial structures. 

For the melt-quench simulations, the interfaces were heated to $600$K, which is above the melting point of Li ($\sim 453\,\mathrm{K}$) but well below that of bulk Cu ($\sim 1358\,\mathrm{K}$). At this temperature, the Li component melts while the Cu surface layers undergo significant disorder due to interfacial mixing with molten Li, producing amorphous interfacial structures representative of realistic Li-Cu interfaces encountered, for example, in lithium-ion battery anodes. The system was then cooled from 600~K to 10~K at a rate of $250$K/ps. The NVT ensemble with the Nose-Hoover thermostat \cite{NVT_1,NVT_2,NVT_3} was used for all the AIMD simulations with a time step of $2$ fs and the energy cutoff fixed as $520$ eV. This interfacial data forms the bulk of our dataset and consists of $11,000$ distinct structures. 

\subsection{\label{sec:dft} DFT Calculations}

We utilized the Vienna ab initio simulation package (VASP)~\cite{VASP_1, VASP_2} with the projector augmented wave \cite{PAW_1, PAW_2} potentials to generate the entire dataset. The restored-regularized strongly constrained and appropriately normed (R\(^2\)SCAN)~\cite{r2scan} exchange correlation functional was used to model the electronic exchange and correlation, with the kinetic energy cutoff fixed as $520$ eV for the plane wave basis. We sampled the Brillouin zone using $\Gamma$ centered Monkhorst-Pack $k$-point meshes having a minimum density of $32$/Å. We allowed for the ionic positions to be relaxed for all configurations without symmetry constraints until the total energies and atomic forces for all structures converged within 10$^{-5}$~eV and $|0.03|$~eV/Å, respectively.

\subsection{\label{sec:allegro+} Allegro enhancements}

The first modification made to the existing Allegro structure was the insertion of an additional MLP layer (see Allegro architecture figure in \cite{allegro_paper}). This MLP consists of two hidden layers, the dimensionality of which (along with the input number of features) consists of three additional hyperparameters for tuning. Moreover, the weights initialization and nonlinearity methods constitute two hyperparameters that we included to the configuration of the model but did not optimize. These MLP layers add some additional depth to the model, which increases the complexity and ultimately leads to better accuracy. 

Allegro$+$MLP serves as a direct classical counterpart to Allegro$+$QDI allowing for a proper comparison between the two models. This comparison is valid since both Allegro$+$MLP and Allegro$+$QDI share the same overall structure. The only difference is that the MLP, including the two hidden layers, is replaced by a QDI layer followed by a fully connected layer, before it passes to the final output MLP of the original Allegro model.

\subsection{\label{sec:hpo} Hyperparameter tuning}

For the hyperparameter optimization of all Allegro, Allegro$+$MLP and Allegro$+$QDI models we used \TQOptimaX{} which evaluated the objective function for $100$ generations and $12$ off-springs, implying a total of $1200$ evaluations for each variant and each dataset. For transparency to computational costs, we report the mean wall-clock time per expensive evaluation on a single T4 GPU at approximately $32.4$ minutes for QM9, $9.6$ minutes for rMD17-aspirin, $6.0$ minutes for rMD17-benzene and $4.8$ minutes for Cu-Li. Multiplying with the number of expensive evaluations of the optimization process, gives an estimate of the total optimization cost per dataset. We optimized Allegro on $6$ different hyperparameters, including the cutoff radius, the polynomial cutoff, batch size, number of layers, dimensionality of the final MLP layer and learning rate. Especially for the Allegro$+$MLP model, we optimized also the dimensionality of the two hidden layers and the number of input features. For Allegro$+$QDI, the additional optimized hyperparameters where the number of input features to the quantum circuit, the dimensionality of the additional MLP and the depth of the QDI.

The models were optimized to the minimize mean absolute error on forces (mean absolute error on energies for QM9) for the validation set, and at the same time the minimization of the inference time of the models on a test set. To visualize and evaluate the optimization, we produced the Pareto fronts for all given datasets Fig.~\ref{fig:pareto}. Unfortunately, we were not able to apply the full Pareto-front optimization on Allegro$+$QDI for the rMD17-aspirin, rMD17-benzene, and QM9 datasets. Consequently, we produced the complete Pareto front for Allegro$+$QDI only on Cu-Li. For the remaining datasets, we transferred the best QDI-specific hyperparameters (QDI depth and QDI input features) found on Cu-Li, while using the dataset-specific Allegro hyperparameters optimized for each respective dataset. We note that this transfer likely represents a lower bound on Allegro$+$QDI performance, since dataset-specific optimization of the QDI hyperparameters could yield further improvements.

\begin{table*}
    \centering
    \begin{tabular}{lcccc}
         $\mathrm{Val}_\mathrm{f_{mae}}$ & Allegro & Allegro \cite{allegro_paper} & Allegro+MLP & Allegro+QDI  \\ \hline
          aspirin & 10.2 & 9.8 & \textbf{9.0} & 9.9\\
          benzene & 0.43 & 0.37 & \textbf{0.28} & 0.30 \\
         CuLi & 45.9 & - & 43.3 & \textbf{37.8} \\ 
    \end{tabular}
    \caption{$\mathrm{Val}_\mathrm{f_{mae}}$ values for best runs in (meV/\AA)}
    \label{tab:forces_results}
\end{table*}

\begin{table*}
    \centering
    \begin{tabular}{lcccc}
         $\mathrm{Val}_\mathrm{e_{mae}}$ & Allegro & Allegro \cite{allegro_paper} & Allegro+MLP & Allegro+QDI  \\ \hline
          aspirin & 3.6 & \textbf{2.3} & 2.6 & 3.2\\
          benzene & 0.21 & 0.20 & 0.21 & \textbf{0.19} \\
         CuLi & 809.0 & - & 698.9 & \textbf{495.7} \\ 
         QM9  & \textbf{10.3} & 13.4 & 21.4 & 23.3 \\
    \end{tabular}
    \caption{$\mathrm{Val}_\mathrm{e_{mae}}$ values for best runs in (meV/atom) for up to 300 epochs on QM9, 100 for CuLi and 3000 for aspirin and benzene.}
    \label{tab:energies_results}
\end{table*}

\section{\label{sec:Results} Results}

\begin{table*}
    \centering
    \begin{tabular}{lccccc}
         & \multicolumn{2}{c}{rMD17-aspirin} & \multicolumn{2}{c}{rMD17-benzene} & QM9 \\
         Model & $\mathrm{F_{MAE}}$ & $\mathrm{E_{MAE}}$ & $\mathrm{F_{MAE}}$ & $\mathrm{E_{MAE}}$ & $\mathrm{E_{MAE}}$ \\
         & (meV/\AA) & (meV/atom) & (meV/\AA) & (meV/atom) & (meV/atom) \\ \hline
         NequIP \cite{nequip_paper} & 8.2 & 2.3 & 0.3 & 0.04 & - \\
         MACE \cite{mace_paper} & 6.6 & 2.2 & 0.3 & 0.4 & - \\
         BOTNet \cite{botnet_paper} & 8.5 & 2.3 & 0.3 & 0.03 & - \\
         Allegro \cite{allegro_paper} & 7.3 & 2.3 & 0.2 & 0.3 & 13.4 \\
         Allegro (ours) & 10.2 & 3.6 & 0.43 & 0.21 & \textbf{10.3} \\
         Allegro+MLP (ours) & \textbf{9.0} & 2.6 & \textbf{0.28} & 0.21 & 21.4 \\
         Allegro+QDI (ours)$^*$ & 9.9 & 3.2 & 0.30 & \textbf{0.19} & 23.3 \\
    \end{tabular}
    \caption{Comparison with state-of-the-art MLIP models on rMD17 and QM9 benchmarks. Values for NequIP, MACE, and BOTNet are taken from Table~1 of Ref.~\cite{mace_paper}, trained on 950 configurations with 50 for validation. $^*$Allegro+QDI results on rMD17 and QM9 use transferred hyperparameters from Cu-Li (see text).}
    \label{tab:sota_comparison}
\end{table*}

By running the multi-objective optimization on all Allegro variants and all datasets we plot the respective Pareto fronts in Fig.~\ref{fig:pareto}. For all datasets, there is a trade-off between accuracy and inference time. For rMD17-aspirin and rMD17-benzene we notice that Allegro$+$MLP can achieve better accuracy but pure Allegro validates a test set faster, while for QM9 it is the opposite. Finally, CuLi dataset in Fig.~\ref{fig:pareto} includes the Allegro$+$QDI Pareto front, which we manually calculated due to technical reasons with the inference time and the optimization library that we use. In principle, it is possible to modify our approach for the QDI variant to make it generally efficient. Nevertheless, for CuLi, Allegro$+$QDI has the overall best accuracy on predicting the forces on the validation set and slightly worse inference  time in comparison with Allegro$+$MLP.

In Tables~\ref{tab:allegro_hpo_params}, ~\ref{tab:allegro_mlp_hpo_params} and ~\ref{tab:allegro_qdi_hpo_params} we include all the best hyperparameters found by \TQOptimaX{} on vanilla Allegro, Allegro+MLP and Allegro+QDI respectively. We notice, large differences in specific hyperparameters (such as the polynomial envelope P and the dimensionality of the edge MLP), between the different variants, which is an indication of a non-significant hyperparameter. Due to the aforementioned technical issue, we optimize Allegro+QDI only on the forces MAE. Specifically for the QM9 dataset, instead of optimizing the Allegro+QDI model we use the optimized hyperparameters on QM9 from Allegro and for the QDI-exclusive hyperparameters, we use the optimized Allegro+QDI parameters from CuLi.

In Tables~\ref{tab:forces_results} and \ref{tab:energies_results} we report the results from training the models on all datasets, using the optimized hyperparameters. In the following supplementary material, we include all the graphs from the training of these runs. We compare all our results with the results reported in the original Allegro paper \cite{allegro_paper}. Since we were unable to replicate the exact configuration file as the one used in \cite{allegro_paper}, we used a common configuration file for all models and datasets and only changed the optimized hyperparameters in each case. In the tables, we denote with bold the most accurate models, which in all cases (for forces) are the variant Allegro+MLP or Allegro+QDI. The Allegro+QDI variant dominates on the CuLi dataset, offering almost $13\%$ increase in performance over Allegro+MLP. 

Table~\ref{tab:sota_comparison} places our results in the context of other state-of-the-art MLIP models. On rMD17-aspirin, our Allegro+MLP variant (9.0~meV/\AA) achieves competitive force MAE, outperforming the original Allegro and BOTNet, though MACE achieves a lower error (6.6~meV/\AA). On rMD17-benzene, Allegro+MLP (0.28~meV/\AA) is competitive with NequIP and MACE. We emphasize that the primary contribution of this work is not to claim state-of-the-art accuracy on these benchmarks, but rather to demonstrate the value of systematic multi-objective hyperparameter optimization and to explore the potential of quantum-classical hybridization within a specific MLIP framework. Direct comparisons should also account for differences in training configurations, data splits, and computational resources.

Regarding energies, we notice a more unstable behavior during training which is justified in such accurate calculations (in the range of a few meV and less than 1~meV for benzene). Since the optimization focuses only on forces and inference time, the energy results do not highlight any model to be strictly superior. 

\section{\label{sec:Discussion} Discussion}

To summarize, we use \TQOptimaX{} to optimize hyperparameters on an Allegro model along with the classical and hybrid variants we built. The optimization targeted two objectives: minimizing the mean absolute error on validation set forces and minimizing inference time. The results indicate that the classical variant Allegro+MLP consistently surpasses vanilla Allegro in force prediction accuracy across all tested datasets. The quantum-classical hybrid Allegro+QDI achieves the best force prediction accuracy on the Cu-Li dataset, where it was fully optimized, offering approximately 13\% improvement over Allegro+MLP. Notably, even with hyperparameters transferred from Cu-Li, Allegro+QDI shows competitive performance on the remaining datasets and achieves the best energy MAE on rMD17-benzene, suggesting that the quantum hybridization benefits are not limited to a single dataset.

For future work, one could explore the capabilities of such state-of-the-art MLIP models (e.g., Allegro), on more complex molecules that are impossible to solve analytically, where such highly accurate, approximation methods can prove significant utility. Based on our work here, we expect the hybridization of such MLIPs to offer additional accuracy benefits. Thus, exploring different ansatzes or involving more classical/quantum layers in the architecture of the model can improve them significantly.

\section*{Data Availability Statement}

The datasets used in this study are publicly available. The rMD17 dataset is available as described in~\cite{rmd17}, and the QM9 dataset in~\cite{qm9-1}. The Cu-Li dataset was generated in collaboration with an industrial partner and is subject to data sharing restrictions. It is available upon reasonable request to the corresponding author for academic research purposes, subject to data sharing agreement.

The machine learning models and code used in this work include NequIP~\cite{nequip_paper}, Allegro~\cite{allegro_paper}, Allegro+MLP~\cite{glatq2025allegro_mlp}, and SAMO-COBRA~\cite{samo2023cobra}, all of which are accessible via their respective GitHub repositories.

\bibliography{main}

\begin{thebibliography}{56}%
\makeatletter
\providecommand \@ifxundefined [1]{%
 \@ifx{#1\undefined}
}%
\providecommand \@ifnum [1]{%
 \ifnum #1\expandafter \@firstoftwo
 \else \expandafter \@secondoftwo
 \fi
}%
\providecommand \@ifx [1]{%
 \ifx #1\expandafter \@firstoftwo
 \else \expandafter \@secondoftwo
 \fi
}%
\providecommand \natexlab [1]{#1}%
\providecommand \enquote  [1]{``#1''}%
\providecommand \bibnamefont  [1]{#1}%
\providecommand \bibfnamefont [1]{#1}%
\providecommand \citenamefont [1]{#1}%
\providecommand \href@noop [0]{\@secondoftwo}%
\providecommand \href [0]{\begingroup \@sanitize@url \@href}%
\providecommand \@href[1]{\@@startlink{#1}\@@href}%
\providecommand \@@href[1]{\endgroup#1\@@endlink}%
\providecommand \@sanitize@url [0]{\catcode `\\12\catcode `\$12\catcode
  `\&12\catcode `\#12\catcode `\^12\catcode `\_12\catcode `\%12\relax}%
\providecommand \@@startlink[1]{}%
\providecommand \@@endlink[0]{}%
\providecommand \url  [0]{\begingroup\@sanitize@url \@url }%
\providecommand \@url [1]{\endgroup\@href {#1}{\urlprefix }}%
\providecommand \urlprefix  [0]{URL }%
\providecommand \Eprint [0]{\href }%
\providecommand \doibase [0]{https://doi.org/}%
\providecommand \selectlanguage [0]{\@gobble}%
\providecommand \bibinfo  [0]{\@secondoftwo}%
\providecommand \bibfield  [0]{\@secondoftwo}%
\providecommand \translation [1]{[#1]}%
\providecommand \BibitemOpen [0]{}%
\providecommand \bibitemStop [0]{}%
\providecommand \bibitemNoStop [0]{.\EOS\space}%
\providecommand \EOS [0]{\spacefactor3000\relax}%
\providecommand \BibitemShut  [1]{\csname bibitem#1\endcsname}%
\let\auto@bib@innerbib\@empty
\bibitem [{\citenamefont {Deringer}\ \emph {et~al.}(2019)\citenamefont
  {Deringer}, \citenamefont {Caro},\ and\ \citenamefont {Cs{\'a}nyi}}]{mlip_1}%
  \BibitemOpen
  \bibfield  {author} {\bibinfo {author} {\bibfnamefont {V.~L.}\ \bibnamefont
  {Deringer}}, \bibinfo {author} {\bibfnamefont {M.~A.}\ \bibnamefont {Caro}},\
  and\ \bibinfo {author} {\bibfnamefont {G.}~\bibnamefont {Cs{\'a}nyi}},\
  }\bibfield  {title} {\bibinfo {title} {Machine learning interatomic
  potentials as emerging tools for materials science},\ }\href@noop {}
  {\bibfield  {journal} {\bibinfo  {journal} {Advanced Materials}\ }\textbf
  {\bibinfo {volume} {31}},\ \bibinfo {pages} {1902765} (\bibinfo {year}
  {2019})}\BibitemShut {NoStop}%
\bibitem [{\citenamefont {Anstine}\ and\ \citenamefont
  {Isayev}(2023)}]{mlip_2}%
  \BibitemOpen
  \bibfield  {author} {\bibinfo {author} {\bibfnamefont {D.~M.}\ \bibnamefont
  {Anstine}}\ and\ \bibinfo {author} {\bibfnamefont {O.}~\bibnamefont
  {Isayev}},\ }\bibfield  {title} {\bibinfo {title} {Machine learning
  interatomic potentials and long-range physics},\ }\href@noop {} {\bibfield
  {journal} {\bibinfo  {journal} {The Journal of Physical Chemistry A}\
  }\textbf {\bibinfo {volume} {127}},\ \bibinfo {pages} {2417} (\bibinfo {year}
  {2023})}\BibitemShut {NoStop}%
\bibitem [{\citenamefont {Mueller}\ \emph {et~al.}(2020)\citenamefont
  {Mueller}, \citenamefont {Hernandez},\ and\ \citenamefont {Wang}}]{mlip_3}%
  \BibitemOpen
  \bibfield  {author} {\bibinfo {author} {\bibfnamefont {T.}~\bibnamefont
  {Mueller}}, \bibinfo {author} {\bibfnamefont {A.}~\bibnamefont {Hernandez}},\
  and\ \bibinfo {author} {\bibfnamefont {C.}~\bibnamefont {Wang}},\ }\bibfield
  {title} {\bibinfo {title} {Machine learning for interatomic potential
  models},\ }\href {https://doi.org/10.1063/1.5126336} {\bibfield  {journal}
  {\bibinfo  {journal} {The Journal of Chemical Physics}\ }\textbf {\bibinfo
  {volume} {152}},\ \bibinfo {pages} {050902} (\bibinfo {year}
  {2020})}\BibitemShut {NoStop}%
\bibitem [{\citenamefont {Argaman}\ and\ \citenamefont {Makov}(2000)}]{dft_1}%
  \BibitemOpen
  \bibfield  {author} {\bibinfo {author} {\bibfnamefont {N.}~\bibnamefont
  {Argaman}}\ and\ \bibinfo {author} {\bibfnamefont {G.}~\bibnamefont
  {Makov}},\ }\bibfield  {title} {\bibinfo {title} {Density functional theory:
  An introduction},\ }\href@noop {} {\bibfield  {journal} {\bibinfo  {journal}
  {American Journal of Physics}\ }\textbf {\bibinfo {volume} {68}},\ \bibinfo
  {pages} {69} (\bibinfo {year} {2000})}\BibitemShut {NoStop}%
\bibitem [{\citenamefont {Orio}\ \emph {et~al.}(2009)\citenamefont {Orio},
  \citenamefont {Pantazis},\ and\ \citenamefont {Neese}}]{dft_2}%
  \BibitemOpen
  \bibfield  {author} {\bibinfo {author} {\bibfnamefont {M.}~\bibnamefont
  {Orio}}, \bibinfo {author} {\bibfnamefont {D.~A.}\ \bibnamefont {Pantazis}},\
  and\ \bibinfo {author} {\bibfnamefont {F.}~\bibnamefont {Neese}},\ }\bibfield
   {title} {\bibinfo {title} {Density functional theory},\ }\href@noop {}
  {\bibfield  {journal} {\bibinfo  {journal} {Photosynthesis Research}\
  }\textbf {\bibinfo {volume} {102}},\ \bibinfo {pages} {443} (\bibinfo {year}
  {2009})}\BibitemShut {NoStop}%
\bibitem [{\citenamefont {Sch{\"u}tt}\ \emph {et~al.}(2018)\citenamefont
  {Sch{\"u}tt}, \citenamefont {Sauceda}, \citenamefont {Kindermans},
  \citenamefont {Tkatchenko},\ and\ \citenamefont {M{\"u}ller}}]{schnet_1}%
  \BibitemOpen
  \bibfield  {author} {\bibinfo {author} {\bibfnamefont {K.~T.}\ \bibnamefont
  {Sch{\"u}tt}}, \bibinfo {author} {\bibfnamefont {H.~E.}\ \bibnamefont
  {Sauceda}}, \bibinfo {author} {\bibfnamefont {P.-J.}\ \bibnamefont
  {Kindermans}}, \bibinfo {author} {\bibfnamefont {A.}~\bibnamefont
  {Tkatchenko}},\ and\ \bibinfo {author} {\bibfnamefont {K.-R.}\ \bibnamefont
  {M{\"u}ller}},\ }\bibfield  {title} {\bibinfo {title} {{SchNet}--a deep
  learning architecture for molecules and materials},\ }\href
  {https://doi.org/10.1063/1.5019779} {\bibfield  {journal} {\bibinfo
  {journal} {The Journal of Chemical Physics}\ }\textbf {\bibinfo {volume}
  {148}},\ \bibinfo {pages} {241722} (\bibinfo {year} {2018})}\BibitemShut
  {NoStop}%
\bibitem [{\citenamefont {Bart{\'o}k}\ \emph {et~al.}(2010)\citenamefont
  {Bart{\'o}k}, \citenamefont {Payne}, \citenamefont {Kondor},\ and\
  \citenamefont {Cs{\'a}nyi}}]{gap}%
  \BibitemOpen
  \bibfield  {author} {\bibinfo {author} {\bibfnamefont {A.~P.}\ \bibnamefont
  {Bart{\'o}k}}, \bibinfo {author} {\bibfnamefont {M.~C.}\ \bibnamefont
  {Payne}}, \bibinfo {author} {\bibfnamefont {R.}~\bibnamefont {Kondor}},\ and\
  \bibinfo {author} {\bibfnamefont {G.}~\bibnamefont {Cs{\'a}nyi}},\ }\bibfield
   {title} {\bibinfo {title} {Gaussian approximation potentials: The accuracy
  of quantum mechanics, without the electrons},\ }\href@noop {} {\bibfield
  {journal} {\bibinfo  {journal} {Physical Review Letters}\ }\textbf {\bibinfo
  {volume} {104}},\ \bibinfo {pages} {136403} (\bibinfo {year}
  {2010})}\BibitemShut {NoStop}%
\bibitem [{\citenamefont {Deringer}\ \emph {et~al.}(2021)\citenamefont
  {Deringer}, \citenamefont {Bart{\'o}k}, \citenamefont {Bernstein},
  \citenamefont {Wilkins}, \citenamefont {Ceriotti},\ and\ \citenamefont
  {Cs{\'a}nyi}}]{soap-gap}%
  \BibitemOpen
  \bibfield  {author} {\bibinfo {author} {\bibfnamefont {V.~L.}\ \bibnamefont
  {Deringer}}, \bibinfo {author} {\bibfnamefont {A.~P.}\ \bibnamefont
  {Bart{\'o}k}}, \bibinfo {author} {\bibfnamefont {N.}~\bibnamefont
  {Bernstein}}, \bibinfo {author} {\bibfnamefont {D.~M.}\ \bibnamefont
  {Wilkins}}, \bibinfo {author} {\bibfnamefont {M.}~\bibnamefont {Ceriotti}},\
  and\ \bibinfo {author} {\bibfnamefont {G.}~\bibnamefont {Cs{\'a}nyi}},\
  }\bibfield  {title} {\bibinfo {title} {Gaussian process regression for
  materials and molecules},\ }\href@noop {} {\bibfield  {journal} {\bibinfo
  {journal} {Chemical Reviews}\ }\textbf {\bibinfo {volume} {121}},\ \bibinfo
  {pages} {10073} (\bibinfo {year} {2021})}\BibitemShut {NoStop}%
\bibitem [{\citenamefont {Drautz}(2019)}]{ACE}%
  \BibitemOpen
  \bibfield  {author} {\bibinfo {author} {\bibfnamefont {R.}~\bibnamefont
  {Drautz}},\ }\bibfield  {title} {\bibinfo {title} {Atomic cluster expansion
  for accurate and transferable interatomic potentials},\ }\href@noop {}
  {\bibfield  {journal} {\bibinfo  {journal} {Physical Review B}\ }\textbf
  {\bibinfo {volume} {99}},\ \bibinfo {pages} {014104} (\bibinfo {year}
  {2019})}\BibitemShut {NoStop}%
\bibitem [{\citenamefont {Lysogorskiy}\ \emph {et~al.}(2021)\citenamefont
  {Lysogorskiy}, \citenamefont {Oord}, \citenamefont {Bochkarev}, \citenamefont
  {Menon}, \citenamefont {Rinaldi}, \citenamefont {Hammerschmidt},
  \citenamefont {Mrovec}, \citenamefont {Thompson}, \citenamefont {Cs{\'a}nyi},
  \citenamefont {Ortner} \emph {et~al.}}]{PACE}%
  \BibitemOpen
  \bibfield  {author} {\bibinfo {author} {\bibfnamefont {Y.}~\bibnamefont
  {Lysogorskiy}}, \bibinfo {author} {\bibfnamefont {C.~v.~d.}\ \bibnamefont
  {Oord}}, \bibinfo {author} {\bibfnamefont {A.}~\bibnamefont {Bochkarev}},
  \bibinfo {author} {\bibfnamefont {S.}~\bibnamefont {Menon}}, \bibinfo
  {author} {\bibfnamefont {M.}~\bibnamefont {Rinaldi}}, \bibinfo {author}
  {\bibfnamefont {T.}~\bibnamefont {Hammerschmidt}}, \bibinfo {author}
  {\bibfnamefont {M.}~\bibnamefont {Mrovec}}, \bibinfo {author} {\bibfnamefont
  {A.}~\bibnamefont {Thompson}}, \bibinfo {author} {\bibfnamefont
  {G.}~\bibnamefont {Cs{\'a}nyi}}, \bibinfo {author} {\bibfnamefont
  {C.}~\bibnamefont {Ortner}}, \emph {et~al.},\ }\bibfield  {title} {\bibinfo
  {title} {Performant implementation of the atomic cluster expansion ({PACE})
  and application to copper and silicon},\ }\href@noop {} {\bibfield  {journal}
  {\bibinfo  {journal} {npj Computational Materials}\ }\textbf {\bibinfo
  {volume} {7}},\ \bibinfo {pages} {97} (\bibinfo {year} {2021})}\BibitemShut
  {NoStop}%
\bibitem [{\citenamefont {Batzner}\ \emph {et~al.}(2022)\citenamefont
  {Batzner}, \citenamefont {Musaelian}, \citenamefont {Sun}, \citenamefont
  {Geiger}, \citenamefont {Mailoa}, \citenamefont {Kornbluth}, \citenamefont
  {Molinari}, \citenamefont {Smidt},\ and\ \citenamefont
  {Kozinsky}}]{nequip_paper}%
  \BibitemOpen
  \bibfield  {author} {\bibinfo {author} {\bibfnamefont {S.}~\bibnamefont
  {Batzner}}, \bibinfo {author} {\bibfnamefont {A.}~\bibnamefont {Musaelian}},
  \bibinfo {author} {\bibfnamefont {L.}~\bibnamefont {Sun}}, \bibinfo {author}
  {\bibfnamefont {M.}~\bibnamefont {Geiger}}, \bibinfo {author} {\bibfnamefont
  {J.~P.}\ \bibnamefont {Mailoa}}, \bibinfo {author} {\bibfnamefont
  {M.}~\bibnamefont {Kornbluth}}, \bibinfo {author} {\bibfnamefont
  {N.}~\bibnamefont {Molinari}}, \bibinfo {author} {\bibfnamefont {T.~E.}\
  \bibnamefont {Smidt}},\ and\ \bibinfo {author} {\bibfnamefont
  {B.}~\bibnamefont {Kozinsky}},\ }\bibfield  {title} {\bibinfo {title}
  {{E(3)}-equivariant graph neural networks for data-efficient and accurate
  interatomic potentials},\ }\href@noop {} {\bibfield  {journal} {\bibinfo
  {journal} {Nature Communications}\ }\textbf {\bibinfo {volume} {13}},\
  \bibinfo {pages} {2453} (\bibinfo {year} {2022})}\BibitemShut {NoStop}%
\bibitem [{\citenamefont {Musaelian}\ \emph
  {et~al.}(2023{\natexlab{a}})\citenamefont {Musaelian}, \citenamefont
  {Batzner}, \citenamefont {Johansson}, \citenamefont {Sun}, \citenamefont
  {Owen}, \citenamefont {Kornbluth},\ and\ \citenamefont
  {Kozinsky}}]{allegro_paper}%
  \BibitemOpen
  \bibfield  {author} {\bibinfo {author} {\bibfnamefont {A.}~\bibnamefont
  {Musaelian}}, \bibinfo {author} {\bibfnamefont {S.}~\bibnamefont {Batzner}},
  \bibinfo {author} {\bibfnamefont {A.}~\bibnamefont {Johansson}}, \bibinfo
  {author} {\bibfnamefont {L.}~\bibnamefont {Sun}}, \bibinfo {author}
  {\bibfnamefont {C.~J.}\ \bibnamefont {Owen}}, \bibinfo {author}
  {\bibfnamefont {M.}~\bibnamefont {Kornbluth}},\ and\ \bibinfo {author}
  {\bibfnamefont {B.}~\bibnamefont {Kozinsky}},\ }\bibfield  {title} {\bibinfo
  {title} {Learning local equivariant representations for large-scale atomistic
  dynamics},\ }\href@noop {} {\bibfield  {journal} {\bibinfo  {journal} {Nature
  Communications}\ }\textbf {\bibinfo {volume} {14}},\ \bibinfo {pages} {579}
  (\bibinfo {year} {2023}{\natexlab{a}})}\BibitemShut {NoStop}%
\bibitem [{\citenamefont {Sagingalieva}\ \emph {et~al.}(2023)\citenamefont
  {Sagingalieva}, \citenamefont {Kordzanganeh}, \citenamefont {Kenbayev},
  \citenamefont {Kosichkina}, \citenamefont {Tomashuk},\ and\ \citenamefont
  {Melnikov}}]{sagingalieva2023hybrid}%
  \BibitemOpen
  \bibfield  {author} {\bibinfo {author} {\bibfnamefont {A.}~\bibnamefont
  {Sagingalieva}}, \bibinfo {author} {\bibfnamefont {M.}~\bibnamefont
  {Kordzanganeh}}, \bibinfo {author} {\bibfnamefont {N.}~\bibnamefont
  {Kenbayev}}, \bibinfo {author} {\bibfnamefont {D.}~\bibnamefont
  {Kosichkina}}, \bibinfo {author} {\bibfnamefont {T.}~\bibnamefont
  {Tomashuk}},\ and\ \bibinfo {author} {\bibfnamefont {A.}~\bibnamefont
  {Melnikov}},\ }\bibfield  {title} {\bibinfo {title} {Hybrid quantum neural
  network for drug response prediction},\ }\href
  {https://doi.org/10.3390/cancers15102705} {\bibfield  {journal} {\bibinfo
  {journal} {Cancers}\ }\textbf {\bibinfo {volume} {15}},\ \bibinfo {pages}
  {2705} (\bibinfo {year} {2023})}\BibitemShut {NoStop}%
\bibitem [{\citenamefont {Sagingalieva}\ \emph
  {et~al.}(2025{\natexlab{a}})\citenamefont {Sagingalieva}, \citenamefont
  {Komornyik}, \citenamefont {Senokosov}, \citenamefont {Joshi}, \citenamefont
  {Mansell}, \citenamefont {Tsurkan}, \citenamefont {Pinto}, \citenamefont
  {Pflitsch},\ and\ \citenamefont {Melnikov}}]{sagingalieva2025photovoltaic}%
  \BibitemOpen
  \bibfield  {author} {\bibinfo {author} {\bibfnamefont {A.}~\bibnamefont
  {Sagingalieva}}, \bibinfo {author} {\bibfnamefont {S.}~\bibnamefont
  {Komornyik}}, \bibinfo {author} {\bibfnamefont {A.}~\bibnamefont
  {Senokosov}}, \bibinfo {author} {\bibfnamefont {A.}~\bibnamefont {Joshi}},
  \bibinfo {author} {\bibfnamefont {C.}~\bibnamefont {Mansell}}, \bibinfo
  {author} {\bibfnamefont {O.}~\bibnamefont {Tsurkan}}, \bibinfo {author}
  {\bibfnamefont {K.}~\bibnamefont {Pinto}}, \bibinfo {author} {\bibfnamefont
  {M.}~\bibnamefont {Pflitsch}},\ and\ \bibinfo {author} {\bibfnamefont
  {A.}~\bibnamefont {Melnikov}},\ }\bibfield  {title} {\bibinfo {title}
  {Photovoltaic power forecasting using quantum machine learning},\ }\href
  {https://doi.org/10.1016/j.solener.2025.114016} {\bibfield  {journal}
  {\bibinfo  {journal} {Solar Energy}\ }\textbf {\bibinfo {volume} {302}},\
  \bibinfo {pages} {114016} (\bibinfo {year} {2025}{\natexlab{a}})}\BibitemShut
  {NoStop}%
\bibitem [{\citenamefont {Lusnig}\ \emph {et~al.}(2024)\citenamefont {Lusnig},
  \citenamefont {Sagingalieva}, \citenamefont {Surmach}, \citenamefont
  {Protasevich}, \citenamefont {Michiu}, \citenamefont {McLoughlin},
  \citenamefont {Mansell}, \citenamefont {de’Petris}, \citenamefont
  {Bonazza}, \citenamefont {Zanconati} \emph {et~al.}}]{lusnig2024hybrid}%
  \BibitemOpen
  \bibfield  {author} {\bibinfo {author} {\bibfnamefont {L.}~\bibnamefont
  {Lusnig}}, \bibinfo {author} {\bibfnamefont {A.}~\bibnamefont
  {Sagingalieva}}, \bibinfo {author} {\bibfnamefont {M.}~\bibnamefont
  {Surmach}}, \bibinfo {author} {\bibfnamefont {T.}~\bibnamefont
  {Protasevich}}, \bibinfo {author} {\bibfnamefont {O.}~\bibnamefont {Michiu}},
  \bibinfo {author} {\bibfnamefont {J.}~\bibnamefont {McLoughlin}}, \bibinfo
  {author} {\bibfnamefont {C.}~\bibnamefont {Mansell}}, \bibinfo {author}
  {\bibfnamefont {G.}~\bibnamefont {de’Petris}}, \bibinfo {author}
  {\bibfnamefont {D.}~\bibnamefont {Bonazza}}, \bibinfo {author} {\bibfnamefont
  {F.}~\bibnamefont {Zanconati}}, \emph {et~al.},\ }\bibfield  {title}
  {\bibinfo {title} {Hybrid quantum image classification and federated learning
  for hepatic steatosis diagnosis},\ }\href@noop {} {\bibfield  {journal}
  {\bibinfo  {journal} {Diagnostics}\ }\textbf {\bibinfo {volume} {14}},\
  \bibinfo {pages} {558} (\bibinfo {year} {2024})}\BibitemShut {NoStop}%
\bibitem [{\citenamefont {Sedykh}\ \emph {et~al.}(2024)\citenamefont {Sedykh},
  \citenamefont {Podapaka}, \citenamefont {Sagingalieva}, \citenamefont
  {Pinto}, \citenamefont {Pflitsch},\ and\ \citenamefont
  {Melnikov}}]{sedykh2024hybrid}%
  \BibitemOpen
  \bibfield  {author} {\bibinfo {author} {\bibfnamefont {A.}~\bibnamefont
  {Sedykh}}, \bibinfo {author} {\bibfnamefont {M.}~\bibnamefont {Podapaka}},
  \bibinfo {author} {\bibfnamefont {A.}~\bibnamefont {Sagingalieva}}, \bibinfo
  {author} {\bibfnamefont {K.}~\bibnamefont {Pinto}}, \bibinfo {author}
  {\bibfnamefont {M.}~\bibnamefont {Pflitsch}},\ and\ \bibinfo {author}
  {\bibfnamefont {A.}~\bibnamefont {Melnikov}},\ }\bibfield  {title} {\bibinfo
  {title} {Hybrid quantum physics-informed neural networks for simulating
  computational fluid dynamics in complex shapes},\ }\href
  {https://doi.org/10.1088/2632-2153/ad43b2} {\bibfield  {journal} {\bibinfo
  {journal} {Machine Learning: Science and Technology}\ }\textbf {\bibinfo
  {volume} {5}},\ \bibinfo {pages} {025045} (\bibinfo {year}
  {2024})}\BibitemShut {NoStop}%
\bibitem [{\citenamefont {Anoshin}\ \emph {et~al.}(2024)\citenamefont
  {Anoshin}, \citenamefont {Sagingalieva}, \citenamefont {Mansell},
  \citenamefont {Zhiganov}, \citenamefont {Shete}, \citenamefont {Pflitsch},\
  and\ \citenamefont {Melnikov}}]{anoshin2024hybrid}%
  \BibitemOpen
  \bibfield  {author} {\bibinfo {author} {\bibfnamefont {M.}~\bibnamefont
  {Anoshin}}, \bibinfo {author} {\bibfnamefont {A.}~\bibnamefont
  {Sagingalieva}}, \bibinfo {author} {\bibfnamefont {C.}~\bibnamefont
  {Mansell}}, \bibinfo {author} {\bibfnamefont {D.}~\bibnamefont {Zhiganov}},
  \bibinfo {author} {\bibfnamefont {V.}~\bibnamefont {Shete}}, \bibinfo
  {author} {\bibfnamefont {M.}~\bibnamefont {Pflitsch}},\ and\ \bibinfo
  {author} {\bibfnamefont {A.}~\bibnamefont {Melnikov}},\ }\bibfield  {title}
  {\bibinfo {title} {Hybrid quantum cycle generative adversarial network for
  small molecule generation},\ }\href
  {https://doi.org/10.1109/TQE.2024.3414264} {\bibfield  {journal} {\bibinfo
  {journal} {IEEE Transactions on Quantum Engineering}\ }\textbf {\bibinfo
  {volume} {5}},\ \bibinfo {pages} {2500514} (\bibinfo {year}
  {2024})}\BibitemShut {NoStop}%
\bibitem [{\citenamefont {Lee}\ \emph {et~al.}(2025)\citenamefont {Lee},
  \citenamefont {Shin}, \citenamefont {Sagingalieva}, \citenamefont
  {Senokosov}, \citenamefont {Anoshin}, \citenamefont {Tripathi}, \citenamefont
  {Pinto},\ and\ \citenamefont {Melnikov}}]{lee2025steel}%
  \BibitemOpen
  \bibfield  {author} {\bibinfo {author} {\bibfnamefont {N.}~\bibnamefont
  {Lee}}, \bibinfo {author} {\bibfnamefont {M.}~\bibnamefont {Shin}}, \bibinfo
  {author} {\bibfnamefont {A.}~\bibnamefont {Sagingalieva}}, \bibinfo {author}
  {\bibfnamefont {A.}~\bibnamefont {Senokosov}}, \bibinfo {author}
  {\bibfnamefont {M.}~\bibnamefont {Anoshin}}, \bibinfo {author} {\bibfnamefont
  {A.~J.}\ \bibnamefont {Tripathi}}, \bibinfo {author} {\bibfnamefont
  {K.}~\bibnamefont {Pinto}},\ and\ \bibinfo {author} {\bibfnamefont
  {A.}~\bibnamefont {Melnikov}},\ }\bibfield  {title} {\bibinfo {title}
  {Predictive control of blast furnace temperature in steelmaking with hybrid
  depth-infused quantum neural networks},\ }\href@noop {} {\bibfield  {journal}
  {\bibinfo  {journal} {arXiv preprint arXiv:2504.12389}\ } (\bibinfo {year}
  {2025})}\BibitemShut {NoStop}%
\bibitem [{\citenamefont {Sagingalieva}\ \emph
  {et~al.}(2025{\natexlab{b}})\citenamefont {Sagingalieva}, \citenamefont
  {Lusnig}, \citenamefont {Cavalli},\ and\ \citenamefont
  {Melnikov}}]{sagingalieva2025hybrid}%
  \BibitemOpen
  \bibfield  {author} {\bibinfo {author} {\bibfnamefont {A.}~\bibnamefont
  {Sagingalieva}}, \bibinfo {author} {\bibfnamefont {L.}~\bibnamefont
  {Lusnig}}, \bibinfo {author} {\bibfnamefont {F.}~\bibnamefont {Cavalli}},\
  and\ \bibinfo {author} {\bibfnamefont {A.}~\bibnamefont {Melnikov}},\
  }\bibfield  {title} {\bibinfo {title} {Hybrid quantum neural networks for
  computer-aided sex diagnosis in forensic and physical anthropology},\ }\href
  {https://doi.org/10.1016/j.imu.2025.101682} {\bibfield  {journal} {\bibinfo
  {journal} {Informatics in Medicine Unlocked}\ }\textbf {\bibinfo {volume}
  {58}},\ \bibinfo {pages} {101682} (\bibinfo {year}
  {2025}{\natexlab{b}})}\BibitemShut {NoStop}%
\bibitem [{\citenamefont {Tsurkan}\ \emph {et~al.}(2025)\citenamefont
  {Tsurkan}, \citenamefont {Konstantinova}, \citenamefont {Sedykh},
  \citenamefont {Senokosov}, \citenamefont {Tarpanov}, \citenamefont {Anoshin},
  \citenamefont {Sagingalieva},\ and\ \citenamefont
  {Melnikov}}]{tsurkan2026hybrid}%
  \BibitemOpen
  \bibfield  {author} {\bibinfo {author} {\bibfnamefont {O.}~\bibnamefont
  {Tsurkan}}, \bibinfo {author} {\bibfnamefont {A.}~\bibnamefont
  {Konstantinova}}, \bibinfo {author} {\bibfnamefont {A.}~\bibnamefont
  {Sedykh}}, \bibinfo {author} {\bibfnamefont {A.}~\bibnamefont {Senokosov}},
  \bibinfo {author} {\bibfnamefont {D.}~\bibnamefont {Tarpanov}}, \bibinfo
  {author} {\bibfnamefont {M.}~\bibnamefont {Anoshin}}, \bibinfo {author}
  {\bibfnamefont {A.}~\bibnamefont {Sagingalieva}},\ and\ \bibinfo {author}
  {\bibfnamefont {A.}~\bibnamefont {Melnikov}},\ }\bibfield  {title} {\bibinfo
  {title} {Hybrid quantum recurrent neural network for remaining useful life
  prediction},\ }\href@noop {} {\bibfield  {journal} {\bibinfo  {journal}
  {arXiv preprint arXiv:2504.20823}\ } (\bibinfo {year} {2025})}\BibitemShut
  {NoStop}%
\bibitem [{\citenamefont {Kurkin}\ \emph {et~al.}(2025)\citenamefont {Kurkin},
  \citenamefont {Hegemann}, \citenamefont {Kordzanganeh},\ and\ \citenamefont
  {Melnikov}}]{kurkin2025forecasting}%
  \BibitemOpen
  \bibfield  {author} {\bibinfo {author} {\bibfnamefont {A.}~\bibnamefont
  {Kurkin}}, \bibinfo {author} {\bibfnamefont {J.}~\bibnamefont {Hegemann}},
  \bibinfo {author} {\bibfnamefont {M.}~\bibnamefont {Kordzanganeh}},\ and\
  \bibinfo {author} {\bibfnamefont {A.}~\bibnamefont {Melnikov}},\ }\bibfield
  {title} {\bibinfo {title} {Forecasting steam mass flow in power plants using
  the parallel hybrid network},\ }\href
  {https://doi.org/10.1016/j.engappai.2025.111912} {\bibfield  {journal}
  {\bibinfo  {journal} {Engineering Applications of Artificial Intelligence}\
  }\textbf {\bibinfo {volume} {160}},\ \bibinfo {pages} {111912} (\bibinfo
  {year} {2025})}\BibitemShut {NoStop}%
\bibitem [{\citenamefont {Lopatkin}\ \emph {et~al.}(2026)\citenamefont
  {Lopatkin}, \citenamefont {Sagingalieva}, \citenamefont {Lusnig},
  \citenamefont {Protasevich}, \citenamefont {Behnke},\ and\ \citenamefont
  {Melnikov}}]{lopatkin2026selector}%
  \BibitemOpen
  \bibfield  {author} {\bibinfo {author} {\bibfnamefont {V.}~\bibnamefont
  {Lopatkin}}, \bibinfo {author} {\bibfnamefont {A.}~\bibnamefont
  {Sagingalieva}}, \bibinfo {author} {\bibfnamefont {L.}~\bibnamefont
  {Lusnig}}, \bibinfo {author} {\bibfnamefont {T.}~\bibnamefont {Protasevich}},
  \bibinfo {author} {\bibfnamefont {B.}~\bibnamefont {Behnke}},\ and\ \bibinfo
  {author} {\bibfnamefont {A.}~\bibnamefont {Melnikov}},\ }\bibfield  {title}
  {\bibinfo {title} {Quantum hybrid feature selector},\ }\href
  {https://doi.org/10.1140/epjqt/s40507-026-00491-1} {\bibfield  {journal}
  {\bibinfo  {journal} {EPJ Quantum Technology}\ }\textbf {\bibinfo {volume}
  {13}},\ \bibinfo {pages} {49} (\bibinfo {year} {2026})}\BibitemShut {NoStop}%
\bibitem [{\citenamefont {de~Winter}\ \emph {et~al.}(2021)\citenamefont
  {de~Winter}, \citenamefont {van Stein},\ and\ \citenamefont
  {B{\"a}ck}}]{samo-cobra}%
  \BibitemOpen
  \bibfield  {author} {\bibinfo {author} {\bibfnamefont {R.}~\bibnamefont
  {de~Winter}}, \bibinfo {author} {\bibfnamefont {B.}~\bibnamefont {van
  Stein}},\ and\ \bibinfo {author} {\bibfnamefont {T.}~\bibnamefont
  {B{\"a}ck}},\ }\bibfield  {title} {\bibinfo {title} {{SAMO-COBRA}: a fast
  surrogate assisted constrained multi-objective optimization algorithm},\ }in\
  \href {https://doi.org/10.1007/978-3-030-72062-9_22} {\emph {\bibinfo
  {booktitle} {International Conference on Evolutionary Multi-Criterion
  Optimization}}},\ \bibinfo {series} {Lecture Notes in Computer Science},
  Vol.\ \bibinfo {volume} {12654}\ (\bibinfo  {publisher} {Springer
  International Publishing},\ \bibinfo {year} {2021})\ pp.\ \bibinfo {pages}
  {270--282}\BibitemShut {NoStop}%
\bibitem [{\citenamefont {Ruddigkeit}\ \emph {et~al.}(2012)\citenamefont
  {Ruddigkeit}, \citenamefont {van Deursen}, \citenamefont {Blum},\ and\
  \citenamefont {Reymond}}]{qm9_1}%
  \BibitemOpen
  \bibfield  {author} {\bibinfo {author} {\bibfnamefont {L.}~\bibnamefont
  {Ruddigkeit}}, \bibinfo {author} {\bibfnamefont {R.}~\bibnamefont {van
  Deursen}}, \bibinfo {author} {\bibfnamefont {L.~C.}\ \bibnamefont {Blum}},\
  and\ \bibinfo {author} {\bibfnamefont {J.-L.}\ \bibnamefont {Reymond}},\
  }\bibfield  {title} {\bibinfo {title} {Enumeration of 166 billion organic
  small molecules in the chemical universe database {GDB-17}},\ }\href@noop {}
  {\bibfield  {journal} {\bibinfo  {journal} {Journal of Chemical Information
  and Modeling}\ }\textbf {\bibinfo {volume} {52}},\ \bibinfo {pages} {2864}
  (\bibinfo {year} {2012})}\BibitemShut {NoStop}%
\bibitem [{\citenamefont {Christensen}\ and\ \citenamefont {von
  Lilienfeld}(2020)}]{rmd17}%
  \BibitemOpen
  \bibfield  {author} {\bibinfo {author} {\bibfnamefont {A.~S.}\ \bibnamefont
  {Christensen}}\ and\ \bibinfo {author} {\bibfnamefont {O.~A.}\ \bibnamefont
  {von Lilienfeld}},\ }\href {https://doi.org/10.6084/m9.figshare.12672038.v3}
  {\bibinfo {title} {Revised {MD17} dataset ({rMD17})}},\ \bibinfo
  {howpublished}
  {\url{https://figshare.com/articles/dataset/Revised_MD17_dataset_rMD17_/12672038}}
  (\bibinfo {year} {2020}),\ \bibinfo {note} {accessed April 2025}\BibitemShut
  {NoStop}%
\bibitem [{\citenamefont {Grisafi}\ \emph {et~al.}(2019)\citenamefont
  {Grisafi}, \citenamefont {Wilkins}, \citenamefont {Willatt},\ and\
  \citenamefont {Ceriotti}}]{equivariance_in_molecules}%
  \BibitemOpen
  \bibfield  {author} {\bibinfo {author} {\bibfnamefont {A.}~\bibnamefont
  {Grisafi}}, \bibinfo {author} {\bibfnamefont {D.~M.}\ \bibnamefont
  {Wilkins}}, \bibinfo {author} {\bibfnamefont {M.~J.}\ \bibnamefont
  {Willatt}},\ and\ \bibinfo {author} {\bibfnamefont {M.}~\bibnamefont
  {Ceriotti}},\ }\bibfield  {title} {\bibinfo {title} {Atomic-scale
  representation and statistical learning of tensorial properties},\ }in\
  \href@noop {} {\emph {\bibinfo {booktitle} {Machine Learning in Chemistry:
  Data-Driven Algorithms, Learning Systems, and Predictions}}}\ (\bibinfo
  {publisher} {ACS Publications},\ \bibinfo {year} {2019})\ pp.\ \bibinfo
  {pages} {1--21}\BibitemShut {NoStop}%
\bibitem [{\citenamefont {Zhang}\ \emph {et~al.}(2018)\citenamefont {Zhang},
  \citenamefont {Han}, \citenamefont {Wang}, \citenamefont {Car},\ and\
  \citenamefont {E}}]{mlip_with_large_data_1}%
  \BibitemOpen
  \bibfield  {author} {\bibinfo {author} {\bibfnamefont {L.}~\bibnamefont
  {Zhang}}, \bibinfo {author} {\bibfnamefont {J.}~\bibnamefont {Han}}, \bibinfo
  {author} {\bibfnamefont {H.}~\bibnamefont {Wang}}, \bibinfo {author}
  {\bibfnamefont {R.}~\bibnamefont {Car}},\ and\ \bibinfo {author}
  {\bibfnamefont {W.}~\bibnamefont {E}},\ }\bibfield  {title} {\bibinfo {title}
  {Deep potential molecular dynamics: a scalable model with the accuracy of
  quantum mechanics},\ }\href@noop {} {\bibfield  {journal} {\bibinfo
  {journal} {Physical Review Letters}\ }\textbf {\bibinfo {volume} {120}},\
  \bibinfo {pages} {143001} (\bibinfo {year} {2018})}\BibitemShut {NoStop}%
\bibitem [{\citenamefont {Smith}\ \emph {et~al.}(2017)\citenamefont {Smith},
  \citenamefont {Isayev},\ and\ \citenamefont
  {Roitberg}}]{mlip_with_large_data_2}%
  \BibitemOpen
  \bibfield  {author} {\bibinfo {author} {\bibfnamefont {J.~S.}\ \bibnamefont
  {Smith}}, \bibinfo {author} {\bibfnamefont {O.}~\bibnamefont {Isayev}},\ and\
  \bibinfo {author} {\bibfnamefont {A.~E.}\ \bibnamefont {Roitberg}},\
  }\bibfield  {title} {\bibinfo {title} {{ANI-1}: an extensible neural network
  potential with {DFT} accuracy at force field computational cost},\
  }\href@noop {} {\bibfield  {journal} {\bibinfo  {journal} {Chemical Science}\
  }\textbf {\bibinfo {volume} {8}},\ \bibinfo {pages} {3192} (\bibinfo {year}
  {2017})}\BibitemShut {NoStop}%
\bibitem [{\citenamefont {Musaelian}\ \emph
  {et~al.}(2023{\natexlab{b}})\citenamefont {Musaelian}, \citenamefont
  {Johansson}, \citenamefont {Batzner},\ and\ \citenamefont
  {Kozinsky}}]{allegro_application}%
  \BibitemOpen
  \bibfield  {author} {\bibinfo {author} {\bibfnamefont {A.}~\bibnamefont
  {Musaelian}}, \bibinfo {author} {\bibfnamefont {A.}~\bibnamefont
  {Johansson}}, \bibinfo {author} {\bibfnamefont {S.}~\bibnamefont {Batzner}},\
  and\ \bibinfo {author} {\bibfnamefont {B.}~\bibnamefont {Kozinsky}},\
  }\bibfield  {title} {\bibinfo {title} {Scaling the leading accuracy of deep
  equivariant models to biomolecular simulations of realistic size},\ }in\
  \href {https://doi.org/10.1145/3581784.3627041} {\emph {\bibinfo {booktitle}
  {Proceedings of the International Conference for High Performance Computing,
  Networking, Storage and Analysis}}}\ (\bibinfo {year} {2023})\ pp.\ \bibinfo
  {pages} {1--12}\BibitemShut {NoStop}%
\bibitem [{\citenamefont {Fu}\ \emph {et~al.}(2023)\citenamefont {Fu},
  \citenamefont {Musaelian}, \citenamefont {Johansson}, \citenamefont
  {Jaakkola},\ and\ \citenamefont {Kozinsky}}]{allegro_application_2}%
  \BibitemOpen
  \bibfield  {author} {\bibinfo {author} {\bibfnamefont {X.}~\bibnamefont
  {Fu}}, \bibinfo {author} {\bibfnamefont {A.}~\bibnamefont {Musaelian}},
  \bibinfo {author} {\bibfnamefont {A.}~\bibnamefont {Johansson}}, \bibinfo
  {author} {\bibfnamefont {T.}~\bibnamefont {Jaakkola}},\ and\ \bibinfo
  {author} {\bibfnamefont {B.}~\bibnamefont {Kozinsky}},\ }\bibfield  {title}
  {\bibinfo {title} {Learning interatomic potentials at multiple scales},\
  }\href@noop {} {\bibfield  {journal} {\bibinfo  {journal} {arXiv preprint
  arXiv:2310.13756}\ } (\bibinfo {year} {2023})}\BibitemShut {NoStop}%
\bibitem [{\citenamefont {Lim}\ and\ \citenamefont
  {Nelson}(2023)}]{butterfly_paper}%
  \BibitemOpen
  \bibfield  {author} {\bibinfo {author} {\bibfnamefont {L.-H.}\ \bibnamefont
  {Lim}}\ and\ \bibinfo {author} {\bibfnamefont {B.~J.}\ \bibnamefont
  {Nelson}},\ }\bibfield  {title} {\bibinfo {title} {What is\ldots an
  equivariant neural network?},\ }\href {https://doi.org/10.1090/noti2666}
  {\bibfield  {journal} {\bibinfo  {journal} {Notices of the American
  Mathematical Society}\ }\textbf {\bibinfo {volume} {70}},\ \bibinfo {pages}
  {619} (\bibinfo {year} {2023})}\BibitemShut {NoStop}%
\bibitem [{\citenamefont {Cohen}\ and\ \citenamefont
  {Welling}(2016)}]{equivariant_cnn}%
  \BibitemOpen
  \bibfield  {author} {\bibinfo {author} {\bibfnamefont {T.}~\bibnamefont
  {Cohen}}\ and\ \bibinfo {author} {\bibfnamefont {M.}~\bibnamefont
  {Welling}},\ }\bibfield  {title} {\bibinfo {title} {Group equivariant
  convolutional networks},\ }in\ \href@noop {} {\emph {\bibinfo {booktitle}
  {International conference on machine learning}}}\ (\bibinfo {organization}
  {PMLR},\ \bibinfo {year} {2016})\ pp.\ \bibinfo {pages}
  {2990--2999}\BibitemShut {NoStop}%
\bibitem [{\citenamefont {Garcia~Satorras}\ \emph {et~al.}(2021)\citenamefont
  {Garcia~Satorras}, \citenamefont {Hoogeboom},\ and\ \citenamefont
  {Welling}}]{e(n)_equivariant_gnns}%
  \BibitemOpen
  \bibfield  {author} {\bibinfo {author} {\bibfnamefont {V.}~\bibnamefont
  {Garcia~Satorras}}, \bibinfo {author} {\bibfnamefont {E.}~\bibnamefont
  {Hoogeboom}},\ and\ \bibinfo {author} {\bibfnamefont {M.}~\bibnamefont
  {Welling}},\ }\bibfield  {title} {\bibinfo {title} {{E(n)} equivariant graph
  neural networks},\ }in\ \href@noop {} {\emph {\bibinfo {booktitle}
  {International conference on machine learning}}}\ (\bibinfo {organization}
  {PMLR},\ \bibinfo {year} {2021})\ pp.\ \bibinfo {pages}
  {9323--9332}\BibitemShut {NoStop}%
\bibitem [{\citenamefont {Navon}\ \emph {et~al.}(2023)\citenamefont {Navon},
  \citenamefont {Shamsian}, \citenamefont {Achituve}, \citenamefont {Fetaya},
  \citenamefont {Chechik},\ and\ \citenamefont
  {Maron}}]{equivariance_weights_relation}%
  \BibitemOpen
  \bibfield  {author} {\bibinfo {author} {\bibfnamefont {A.}~\bibnamefont
  {Navon}}, \bibinfo {author} {\bibfnamefont {A.}~\bibnamefont {Shamsian}},
  \bibinfo {author} {\bibfnamefont {I.}~\bibnamefont {Achituve}}, \bibinfo
  {author} {\bibfnamefont {E.}~\bibnamefont {Fetaya}}, \bibinfo {author}
  {\bibfnamefont {G.}~\bibnamefont {Chechik}},\ and\ \bibinfo {author}
  {\bibfnamefont {H.}~\bibnamefont {Maron}},\ }\bibfield  {title} {\bibinfo
  {title} {Equivariant architectures for learning in deep weight spaces},\ }in\
  \href@noop {} {\emph {\bibinfo {booktitle} {International Conference on
  Machine Learning}}}\ (\bibinfo {organization} {PMLR},\ \bibinfo {year}
  {2023})\ pp.\ \bibinfo {pages} {25790--25816}\BibitemShut {NoStop}%
\bibitem [{\citenamefont {Ong}\ \emph {et~al.}(2013)\citenamefont {Ong},
  \citenamefont {Richards}, \citenamefont {Jain}, \citenamefont {Hautier},
  \citenamefont {Kocher}, \citenamefont {Cholia}, \citenamefont {Gunter},
  \citenamefont {Chevrier}, \citenamefont {Persson},\ and\ \citenamefont
  {Ceder}}]{pyamtgen}%
  \BibitemOpen
  \bibfield  {author} {\bibinfo {author} {\bibfnamefont {S.~P.}\ \bibnamefont
  {Ong}}, \bibinfo {author} {\bibfnamefont {W.~D.}\ \bibnamefont {Richards}},
  \bibinfo {author} {\bibfnamefont {A.}~\bibnamefont {Jain}}, \bibinfo {author}
  {\bibfnamefont {G.}~\bibnamefont {Hautier}}, \bibinfo {author} {\bibfnamefont
  {M.}~\bibnamefont {Kocher}}, \bibinfo {author} {\bibfnamefont
  {S.}~\bibnamefont {Cholia}}, \bibinfo {author} {\bibfnamefont
  {D.}~\bibnamefont {Gunter}}, \bibinfo {author} {\bibfnamefont {V.~L.}\
  \bibnamefont {Chevrier}}, \bibinfo {author} {\bibfnamefont {K.~A.}\
  \bibnamefont {Persson}},\ and\ \bibinfo {author} {\bibfnamefont
  {G.}~\bibnamefont {Ceder}},\ }\bibfield  {title} {\bibinfo {title} {Python
  materials genomics (pymatgen): A robust, open-source python library for
  materials analysis},\ }\href
  {https://doi.org/10.1016/j.commatsci.2012.10.028} {\bibfield  {journal}
  {\bibinfo  {journal} {Computational Materials Science}\ }\textbf {\bibinfo
  {volume} {68}},\ \bibinfo {pages} {314} (\bibinfo {year} {2013})}\BibitemShut
  {NoStop}%
\bibitem [{\citenamefont {Zur}\ and\ \citenamefont
  {McGill}(1984)}]{Zur_Mcgill}%
  \BibitemOpen
  \bibfield  {author} {\bibinfo {author} {\bibfnamefont {A.}~\bibnamefont
  {Zur}}\ and\ \bibinfo {author} {\bibfnamefont {T.~C.}\ \bibnamefont
  {McGill}},\ }\bibfield  {title} {\bibinfo {title} {Lattice match: An
  application to heteroepitaxy},\ }\href {https://doi.org/10.1063/1.333084}
  {\bibfield  {journal} {\bibinfo  {journal} {Journal of Applied Physics}\
  }\textbf {\bibinfo {volume} {55}},\ \bibinfo {pages} {378} (\bibinfo {year}
  {1984})}\BibitemShut {NoStop}%
\bibitem [{\citenamefont {Hoover}(1985)}]{NVT_1}%
  \BibitemOpen
  \bibfield  {author} {\bibinfo {author} {\bibfnamefont {W.~G.}\ \bibnamefont
  {Hoover}},\ }\bibfield  {title} {\bibinfo {title} {Canonical dynamics:
  Equilibrium phase-space distributions},\ }\href
  {https://doi.org/10.1103/PhysRevA.31.1695} {\bibfield  {journal} {\bibinfo
  {journal} {Physical Review A}\ }\textbf {\bibinfo {volume} {31}},\ \bibinfo
  {pages} {1695} (\bibinfo {year} {1985})}\BibitemShut {NoStop}%
\bibitem [{\citenamefont {Nosé}(1991)}]{NVT_2}%
  \BibitemOpen
  \bibfield  {author} {\bibinfo {author} {\bibfnamefont {S.}~\bibnamefont
  {Nosé}},\ }\bibfield  {title} {\bibinfo {title} {Constant temperature
  molecular dynamics methods},\ }\href {https://doi.org/10.1143/PTPS.103.1}
  {\bibfield  {journal} {\bibinfo  {journal} {Progress of Theoretical Physics
  Supplement}\ }\textbf {\bibinfo {volume} {103}},\ \bibinfo {pages} {1}
  (\bibinfo {year} {1991})}\BibitemShut {NoStop}%
\bibitem [{\citenamefont {Nosé}(1984)}]{NVT_3}%
  \BibitemOpen
  \bibfield  {author} {\bibinfo {author} {\bibfnamefont {S.}~\bibnamefont
  {Nosé}},\ }\bibfield  {title} {\bibinfo {title} {A unified formulation of
  the constant temperature molecular dynamics methods},\ }\href
  {https://doi.org/10.1063/1.447334} {\bibfield  {journal} {\bibinfo  {journal}
  {The Journal of Chemical Physics}\ }\textbf {\bibinfo {volume} {81}},\
  \bibinfo {pages} {511} (\bibinfo {year} {1984})}\BibitemShut {NoStop}%
\bibitem [{\citenamefont {Kresse}\ and\ \citenamefont
  {Furthmüller}(1996)}]{VASP_1}%
  \BibitemOpen
  \bibfield  {author} {\bibinfo {author} {\bibfnamefont {G.}~\bibnamefont
  {Kresse}}\ and\ \bibinfo {author} {\bibfnamefont {J.}~\bibnamefont
  {Furthmüller}},\ }\bibfield  {title} {\bibinfo {title} {Efficiency of
  ab-initio total energy calculations for metals and semiconductors using a
  plane-wave basis set},\ }\href {https://doi.org/10.1016/0927-0256(96)00008-0}
  {\bibfield  {journal} {\bibinfo  {journal} {Computational Materials Science}\
  }\textbf {\bibinfo {volume} {6}},\ \bibinfo {pages} {15} (\bibinfo {year}
  {1996})}\BibitemShut {NoStop}%
\bibitem [{\citenamefont {Kresse}\ and\ \citenamefont
  {Furthm\"uller}(1996)}]{VASP_2}%
  \BibitemOpen
  \bibfield  {author} {\bibinfo {author} {\bibfnamefont {G.}~\bibnamefont
  {Kresse}}\ and\ \bibinfo {author} {\bibfnamefont {J.}~\bibnamefont
  {Furthm\"uller}},\ }\bibfield  {title} {\bibinfo {title} {Efficient iterative
  schemes for ab initio total-energy calculations using a plane-wave basis
  set},\ }\href {https://doi.org/10.1103/PhysRevB.54.11169} {\bibfield
  {journal} {\bibinfo  {journal} {Physical Review B}\ }\textbf {\bibinfo
  {volume} {54}},\ \bibinfo {pages} {11169} (\bibinfo {year}
  {1996})}\BibitemShut {NoStop}%
\bibitem [{\citenamefont {Kresse}\ and\ \citenamefont {Joubert}(1999)}]{PAW_1}%
  \BibitemOpen
  \bibfield  {author} {\bibinfo {author} {\bibfnamefont {G.}~\bibnamefont
  {Kresse}}\ and\ \bibinfo {author} {\bibfnamefont {D.}~\bibnamefont
  {Joubert}},\ }\bibfield  {title} {\bibinfo {title} {From ultrasoft
  pseudopotentials to the projector augmented-wave method},\ }\href
  {https://doi.org/10.1103/PhysRevB.59.1758} {\bibfield  {journal} {\bibinfo
  {journal} {Physical Review B}\ }\textbf {\bibinfo {volume} {59}},\ \bibinfo
  {pages} {1758} (\bibinfo {year} {1999})}\BibitemShut {NoStop}%
\bibitem [{\citenamefont {Perdew}\ \emph {et~al.}(1996)\citenamefont {Perdew},
  \citenamefont {Burke},\ and\ \citenamefont {Ernzerhof}}]{PAW_2}%
  \BibitemOpen
  \bibfield  {author} {\bibinfo {author} {\bibfnamefont {J.~P.}\ \bibnamefont
  {Perdew}}, \bibinfo {author} {\bibfnamefont {K.}~\bibnamefont {Burke}},\ and\
  \bibinfo {author} {\bibfnamefont {M.}~\bibnamefont {Ernzerhof}},\ }\bibfield
  {title} {\bibinfo {title} {Generalized gradient approximation made simple},\
  }\href {https://doi.org/10.1103/PhysRevLett.77.3865} {\bibfield  {journal}
  {\bibinfo  {journal} {Physical Review Letters}\ }\textbf {\bibinfo {volume}
  {77}},\ \bibinfo {pages} {3865} (\bibinfo {year} {1996})}\BibitemShut
  {NoStop}%
\bibitem [{\citenamefont {Furness}\ \emph {et~al.}(2020)\citenamefont
  {Furness}, \citenamefont {Kaplan}, \citenamefont {Ning}, \citenamefont
  {Perdew},\ and\ \citenamefont {Sun}}]{r2scan}%
  \BibitemOpen
  \bibfield  {author} {\bibinfo {author} {\bibfnamefont {J.~W.}\ \bibnamefont
  {Furness}}, \bibinfo {author} {\bibfnamefont {A.~D.}\ \bibnamefont {Kaplan}},
  \bibinfo {author} {\bibfnamefont {J.}~\bibnamefont {Ning}}, \bibinfo {author}
  {\bibfnamefont {J.~P.}\ \bibnamefont {Perdew}},\ and\ \bibinfo {author}
  {\bibfnamefont {J.}~\bibnamefont {Sun}},\ }\bibfield  {title} {\bibinfo
  {title} {Accurate and numerically efficient {r2SCAN} meta-generalized
  gradient approximation},\ }\href
  {https://doi.org/10.1021/acs.jpclett.0c02405} {\bibfield  {journal} {\bibinfo
   {journal} {The Journal of Physical Chemistry Letters}\ }\textbf {\bibinfo
  {volume} {11}},\ \bibinfo {pages} {8208} (\bibinfo {year}
  {2020})}\BibitemShut {NoStop}%
\bibitem [{\citenamefont {Batatia}\ \emph {et~al.}(2022)\citenamefont
  {Batatia}, \citenamefont {Kov{\'a}cs}, \citenamefont {Simm}, \citenamefont
  {Ortner},\ and\ \citenamefont {Cs{\'a}nyi}}]{mace_paper}%
  \BibitemOpen
  \bibfield  {author} {\bibinfo {author} {\bibfnamefont {I.}~\bibnamefont
  {Batatia}}, \bibinfo {author} {\bibfnamefont {D.~P.}\ \bibnamefont
  {Kov{\'a}cs}}, \bibinfo {author} {\bibfnamefont {G.~N.~C.}\ \bibnamefont
  {Simm}}, \bibinfo {author} {\bibfnamefont {C.}~\bibnamefont {Ortner}},\ and\
  \bibinfo {author} {\bibfnamefont {G.}~\bibnamefont {Cs{\'a}nyi}},\ }\bibfield
   {title} {\bibinfo {title} {{MACE}: Higher order equivariant message passing
  neural networks for fast and accurate force fields},\ }in\ \href@noop {}
  {\emph {\bibinfo {booktitle} {Advances in Neural Information Processing
  Systems}}},\ Vol.~\bibinfo {volume} {35}\ (\bibinfo  {publisher} {Curran
  Associates, Inc.},\ \bibinfo {year} {2022})\ pp.\ \bibinfo {pages}
  {11423--11436}\BibitemShut {NoStop}%
\bibitem [{\citenamefont {Batatia}\ \emph {et~al.}(2025)\citenamefont
  {Batatia}, \citenamefont {Batzner}, \citenamefont {Kov{\'a}cs}, \citenamefont
  {Musaelian}, \citenamefont {Simm}, \citenamefont {Drautz}, \citenamefont
  {Ortner}, \citenamefont {Kozinsky},\ and\ \citenamefont
  {Cs{\'a}nyi}}]{botnet_paper}%
  \BibitemOpen
  \bibfield  {author} {\bibinfo {author} {\bibfnamefont {I.}~\bibnamefont
  {Batatia}}, \bibinfo {author} {\bibfnamefont {S.}~\bibnamefont {Batzner}},
  \bibinfo {author} {\bibfnamefont {D.~P.}\ \bibnamefont {Kov{\'a}cs}},
  \bibinfo {author} {\bibfnamefont {A.}~\bibnamefont {Musaelian}}, \bibinfo
  {author} {\bibfnamefont {G.~N.~C.}\ \bibnamefont {Simm}}, \bibinfo {author}
  {\bibfnamefont {R.}~\bibnamefont {Drautz}}, \bibinfo {author} {\bibfnamefont
  {C.}~\bibnamefont {Ortner}}, \bibinfo {author} {\bibfnamefont
  {B.}~\bibnamefont {Kozinsky}},\ and\ \bibinfo {author} {\bibfnamefont
  {G.}~\bibnamefont {Cs{\'a}nyi}},\ }\bibfield  {title} {\bibinfo {title} {The
  design space of {E}(3)-equivariant atom-centred interatomic potentials},\
  }\href {https://doi.org/10.1038/s42256-024-00956-x} {\bibfield  {journal}
  {\bibinfo  {journal} {Nature Machine Intelligence}\ }\textbf {\bibinfo
  {volume} {7}},\ \bibinfo {pages} {56} (\bibinfo {year} {2025})}\BibitemShut
  {NoStop}%
\bibitem [{\citenamefont {Ramakrishnan}\ \emph {et~al.}(2014)\citenamefont
  {Ramakrishnan}, \citenamefont {Dral}, \citenamefont {Rupp},\ and\
  \citenamefont {von Lilienfeld}}]{qm9-1}%
  \BibitemOpen
  \bibfield  {author} {\bibinfo {author} {\bibfnamefont {R.}~\bibnamefont
  {Ramakrishnan}}, \bibinfo {author} {\bibfnamefont {P.~O.}\ \bibnamefont
  {Dral}}, \bibinfo {author} {\bibfnamefont {M.}~\bibnamefont {Rupp}},\ and\
  \bibinfo {author} {\bibfnamefont {O.~A.}\ \bibnamefont {von Lilienfeld}},\
  }\bibfield  {title} {\bibinfo {title} {Quantum chemistry structures and
  properties of 134 kilo molecules},\ }\href
  {https://doi.org/10.1038/sdata.2014.22} {\bibfield  {journal} {\bibinfo
  {journal} {Scientific Data}\ }\textbf {\bibinfo {volume} {1}},\ \bibinfo
  {pages} {140022} (\bibinfo {year} {2014})}\BibitemShut {NoStop}%
\bibitem [{gla(2025)}]{glatq2025allegro_mlp}%
  \BibitemOpen
  \href@noop {} {\bibinfo {title} {{Allegro+MLP}: A modification of {Allegro}
  to include {MLP} components}},\ \bibinfo {howpublished}
  {\url{https://github.com/glatq/allegro}} (\bibinfo {year} {2025}),\ \bibinfo
  {note} {{GitHub} repository}\BibitemShut {NoStop}%
\bibitem [{\citenamefont {de~Winter}(2021)}]{samo2023cobra}%
  \BibitemOpen
  \bibfield  {author} {\bibinfo {author} {\bibfnamefont {R.}~\bibnamefont
  {de~Winter}},\ }\href@noop {} {\bibinfo {title} {{SAMO-COBRA}: A fast
  surrogate-assisted constrained multi-objective optimization algorithm}},\
  \bibinfo {howpublished} {\url{https://github.com/RoydeZomer/SAMO-COBRA}}
  (\bibinfo {year} {2021}),\ \bibinfo {note} {{GitHub} repository}\BibitemShut
  {NoStop}%
\bibitem [{\citenamefont {Coecke}\ and\ \citenamefont
  {Duncan}(2008)}]{duncan2022quantum}%
  \BibitemOpen
  \bibfield  {author} {\bibinfo {author} {\bibfnamefont {B.}~\bibnamefont
  {Coecke}}\ and\ \bibinfo {author} {\bibfnamefont {R.}~\bibnamefont
  {Duncan}},\ }\bibfield  {title} {\bibinfo {title} {Interacting quantum
  observables},\ }in\ \href {https://doi.org/10.1007/978-3-540-70583-3_25}
  {\emph {\bibinfo {booktitle} {Automata, Languages and Programming (ICALP
  2008)}}},\ \bibinfo {series} {Lecture Notes in Computer Science}, Vol.\
  \bibinfo {volume} {5126}\ (\bibinfo  {publisher} {Springer},\ \bibinfo {year}
  {2008})\ pp.\ \bibinfo {pages} {298--310}\BibitemShut {NoStop}%
\bibitem [{\citenamefont {van~de Wetering}(2020)}]{wetering2020zx}%
  \BibitemOpen
  \bibfield  {author} {\bibinfo {author} {\bibfnamefont {J.}~\bibnamefont
  {van~de Wetering}},\ }\bibfield  {title} {\bibinfo {title} {{ZX}-calculus for
  the working quantum computer scientist},\ }\href@noop {} {\bibfield
  {journal} {\bibinfo  {journal} {arXiv preprint arXiv:2012.13966}\ } (\bibinfo
  {year} {2020})}\BibitemShut {NoStop}%
\bibitem [{\citenamefont {Abbas}\ \emph {et~al.}(2021)\citenamefont {Abbas},
  \citenamefont {Sutter}, \citenamefont {Zoufal}, \citenamefont {Lucchi},
  \citenamefont {Figalli},\ and\ \citenamefont {Woerner}}]{abbas2020power}%
  \BibitemOpen
  \bibfield  {author} {\bibinfo {author} {\bibfnamefont {A.}~\bibnamefont
  {Abbas}}, \bibinfo {author} {\bibfnamefont {D.}~\bibnamefont {Sutter}},
  \bibinfo {author} {\bibfnamefont {C.}~\bibnamefont {Zoufal}}, \bibinfo
  {author} {\bibfnamefont {A.}~\bibnamefont {Lucchi}}, \bibinfo {author}
  {\bibfnamefont {A.}~\bibnamefont {Figalli}},\ and\ \bibinfo {author}
  {\bibfnamefont {S.}~\bibnamefont {Woerner}},\ }\bibfield  {title} {\bibinfo
  {title} {The power of quantum neural networks},\ }\href
  {https://doi.org/10.1038/s43588-021-00084-1} {\bibfield  {journal} {\bibinfo
  {journal} {Nature Computational Science}\ }\textbf {\bibinfo {volume} {1}},\
  \bibinfo {pages} {403} (\bibinfo {year} {2021})}\BibitemShut {NoStop}%
\bibitem [{\citenamefont {Amari}(1998)}]{amari1998gradient}%
  \BibitemOpen
  \bibfield  {author} {\bibinfo {author} {\bibfnamefont {S.-i.}\ \bibnamefont
  {Amari}},\ }\bibfield  {title} {\bibinfo {title} {Natural gradient works
  efficiently in learning},\ }\href@noop {} {\bibfield  {journal} {\bibinfo
  {journal} {Neural Computation}\ }\textbf {\bibinfo {volume} {10}},\ \bibinfo
  {pages} {251} (\bibinfo {year} {1998})}\BibitemShut {NoStop}%
\bibitem [{\citenamefont {McClean}\ \emph {et~al.}(2018)\citenamefont
  {McClean}, \citenamefont {Boixo}, \citenamefont {Smelyanskiy}, \citenamefont
  {Babbush},\ and\ \citenamefont {Neven}}]{mcclean2018barren}%
  \BibitemOpen
  \bibfield  {author} {\bibinfo {author} {\bibfnamefont {J.~R.}\ \bibnamefont
  {McClean}}, \bibinfo {author} {\bibfnamefont {S.}~\bibnamefont {Boixo}},
  \bibinfo {author} {\bibfnamefont {V.~N.}\ \bibnamefont {Smelyanskiy}},
  \bibinfo {author} {\bibfnamefont {R.}~\bibnamefont {Babbush}},\ and\ \bibinfo
  {author} {\bibfnamefont {H.}~\bibnamefont {Neven}},\ }\bibfield  {title}
  {\bibinfo {title} {Barren plateaus in quantum neural network training
  landscapes},\ }\href {https://doi.org/10.1038/s41467-018-07090-4} {\bibfield
  {journal} {\bibinfo  {journal} {Nature Communications}\ }\textbf {\bibinfo
  {volume} {9}},\ \bibinfo {pages} {4812} (\bibinfo {year} {2018})}\BibitemShut
  {NoStop}%
\bibitem [{\citenamefont {Larocca}\ \emph {et~al.}(2023)\citenamefont
  {Larocca}, \citenamefont {Ju}, \citenamefont {Garc{\'\i}a-Mart{\'\i}n},
  \citenamefont {Coles},\ and\ \citenamefont {Cerezo}}]{larocca2021overparam}%
  \BibitemOpen
  \bibfield  {author} {\bibinfo {author} {\bibfnamefont {M.}~\bibnamefont
  {Larocca}}, \bibinfo {author} {\bibfnamefont {N.}~\bibnamefont {Ju}},
  \bibinfo {author} {\bibfnamefont {D.}~\bibnamefont
  {Garc{\'\i}a-Mart{\'\i}n}}, \bibinfo {author} {\bibfnamefont {P.~J.}\
  \bibnamefont {Coles}},\ and\ \bibinfo {author} {\bibfnamefont
  {M.}~\bibnamefont {Cerezo}},\ }\bibfield  {title} {\bibinfo {title} {Theory
  of overparametrization in quantum neural networks},\ }\href@noop {}
  {\bibfield  {journal} {\bibinfo  {journal} {Nature Computational Science}\
  }\textbf {\bibinfo {volume} {3}},\ \bibinfo {pages} {542} (\bibinfo {year}
  {2023})}\BibitemShut {NoStop}%
\bibitem [{\citenamefont {Luxenberg}\ and\ \citenamefont
  {Boyd}(2024)}]{time_EMA}%
  \BibitemOpen
  \bibfield  {author} {\bibinfo {author} {\bibfnamefont {E.}~\bibnamefont
  {Luxenberg}}\ and\ \bibinfo {author} {\bibfnamefont {S.}~\bibnamefont
  {Boyd}},\ }\bibfield  {title} {\bibinfo {title} {Exponentially weighted
  moving models},\ }\href@noop {} {\bibfield  {journal} {\bibinfo  {journal}
  {arXiv preprint arXiv:2404.08136}\ } (\bibinfo {year} {2024})}\BibitemShut
  {NoStop}%
\end{thebibliography}%

\clearpage

\section*{Supplementary material}

\begin{figure*}[!htbp]
    \centering
    \includegraphics[width=\textwidth]{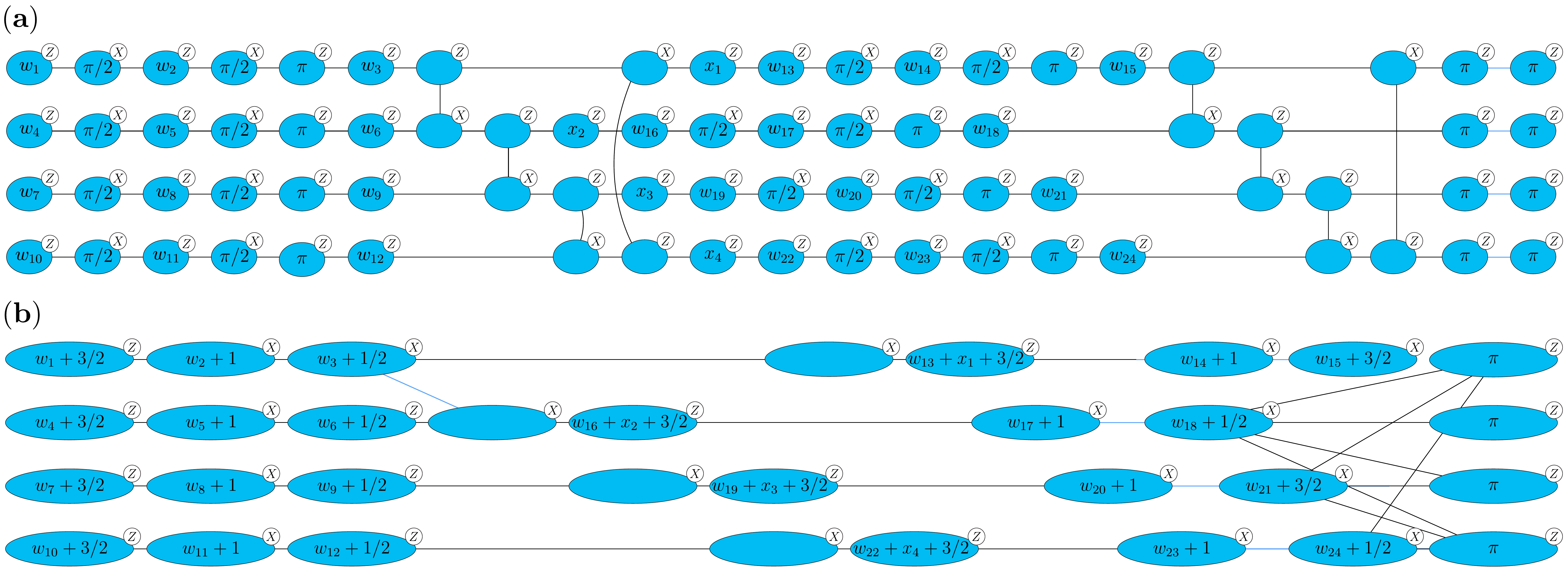}
    \caption{\textbf{(a)} Circuit's ZX graph with original parameters. \textbf{(b)} Circuit's ZX graph after simplification.}
    \label{fig:ZX}
\end{figure*}

\subsection{Quantum Circuits Analysis}\label{sec:appendix_circs}

In this section, we analyze the quantum circuit used in this paper. There are three metrics for evaluation:
\begin{itemize}
\item Redundancy analysis with ZX calculus
\item Trainability analysis with Fisher information
\item Expressivity analysis with Fourier series
\end{itemize}.

\subsubsection{Redundancy analysis with ZX calculus}\label{sec:appendix_ZX}
ZX calculus is a graphical language used to simplify quantum circuits using specific notation, where the circuit is transformed into a sequence of nodes, called "spiders", and edges to connect them. After this ZX graph can be simplified ~\cite{duncan2022quantum} using language's set of rules, thus taking out redundant elements ~\cite{wetering2020zx}. After transformation, we can compare the initial version with the reduced one, which represents a more optimal configuration. Using the number of parameters in the first and second circuits, the redundancy can be evaluated as the margin of initial parameters left after simplification. The more parameters are left unchanged, the better architecture the circuit has in the first place. A architecture deemed unable to be simplified in this manner is classified as ZX-irreducible.

The main changes in the circuit shown in Fig.~\ref{fig:ZX}(a-b) consist of the rearrangement of certain weights after their simplification. During the optimization process, $24$ out of $24$ parameters, or $100\%$, were preserved, showing the circuit's high quality and lack of redundant elements.

Using ZX calculus, it can be clearly said that a circuit does not need major structural changes and has no redundant elements in it. However, other metrics should be used to provide a full analysis of its capabilities.

\subsubsection{Trainability analysis with Fisher information}\label{sec:appendix_fim}

A supervised machine learning task can be framed as constructing a hypothesis model \( h_\theta (\hat{x}) \) based on a labeled dataset \( (x, y) \in X \times Y \) to approximate the natural data distribution \( f(x) \). Given a subset \( S \) of labeled samples from this distribution, we aim to optimize our model to accurately approximate \( f(\hat{x}) \). This involves maximizing the probability of the model, with parameters \( \theta \), predicting the correct label \( y \) for each data point \( x \). The conditional probability required for this task can be expressed as \( P(y|x, \theta) \). By assuming a uniform distribution over \( X \), we leverage the joint probability \( P(y, x | \theta) \) to enhance accuracy, with this probability distribution calculable for any parameter value \( \theta \) given a data point \( x_i \). This joint probability can be represented as an \( N \)-dimensional manifold, where \( N \) is the number of trainable parameters, \( N = |\theta| \).

The Fisher information matrix (FIM) $F(\theta)$~\cite{abbas2020power, amari1998gradient} is a metric over this manifold.

\begin{figure*}[!ht]
    \centering
    \includegraphics[width=1\textwidth]{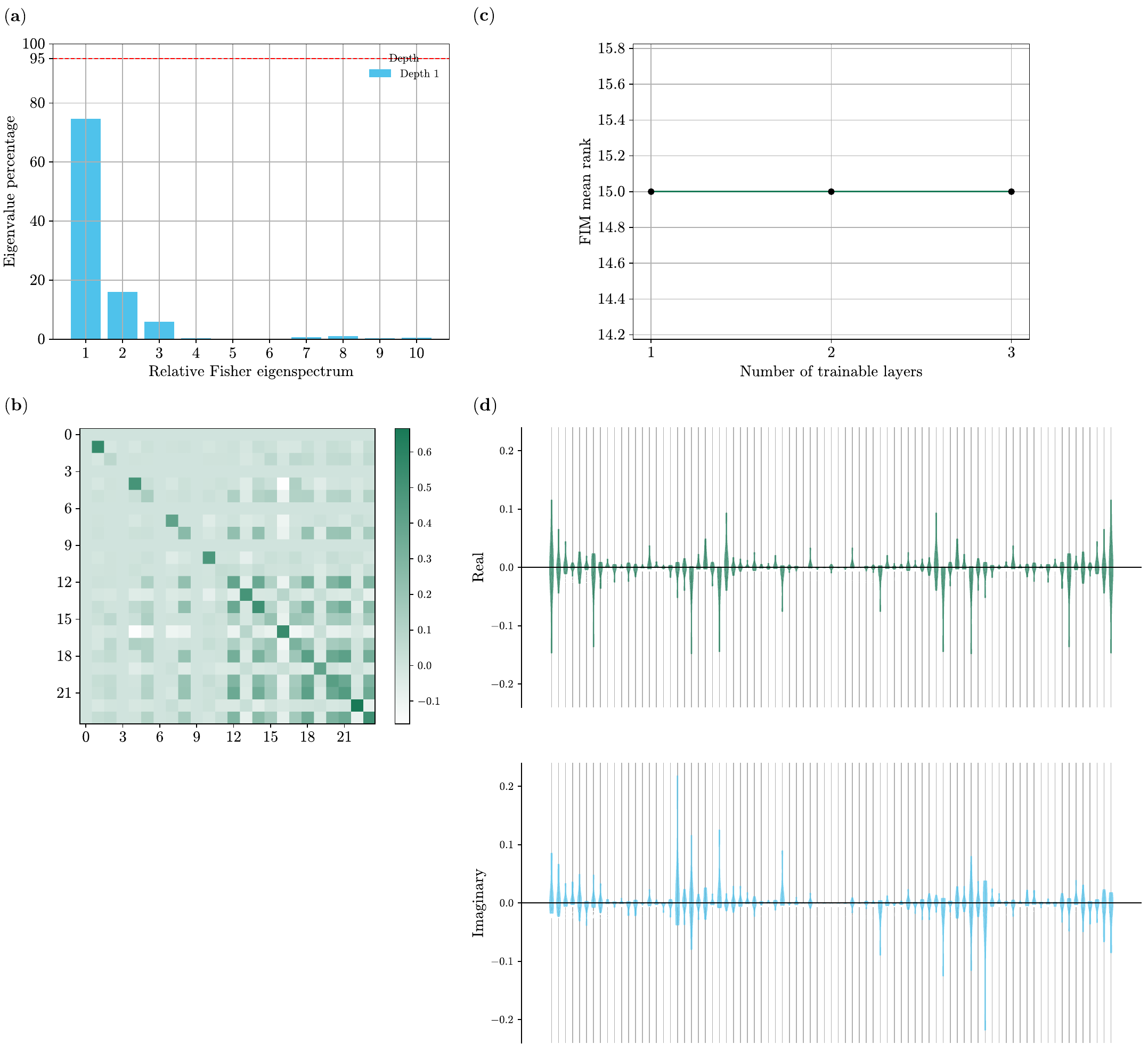}
    \caption{\textbf{(a)} The histogram of the Fisher eigenspectrum. There are around 78\% of parameters with near-zero eigenvalues, but this margin does not go higher than 95\% trainability threshold. Other parameters have great impact, contributing to circuit's trainability. \textbf{(b)} Average Fisher Information Matrix. The diagonal elements show parameters having mostly high gradient values, showing their impact on the circuit. Meanwhile non-diagonal ones are almost non-existent, indicating parameters' independence. Also there is no evident single-parameter dominance, showing high trainability. \textbf{(c)} Average FIM rank. There are three layers presented, although only one is used in the original circuit. This clearly shows how rank goes on the plateau, which means the model is in the overparameterization mode. Thus, no extra layers are needed as the circuit already has a sufficient number of parameters. \textbf{(d)} Real and imaginary part of Fourier coefficients}
    \label{fig:QC}
\end{figure*}

\begin{equation}
    F(\theta)=\mathbb{E}_{\left\{x_i,y_i\right\}}[\nabla_\theta \log(P) \nabla_\theta \log(P)^T]
\end{equation}

To proceed, we diagonalize the metric to obtain a locally Euclidean tangential basis, where the diagonal elements represent the squared gradients of our joint probability within this basis. These elements correspond to the eigenvalues of the Fisher information matrix. This process is crucial for identifying and mitigating the barren plateau problem, which manifests as vanishing gradients in quantum neural networks with a high number of qubits.

As demonstrated in Ref.~\cite{mcclean2018barren}, the expectation values of gradients approach zero, and their variance diminishes exponentially as the number of qubits increases. This effect arises when gradients predominantly converge near zero, leading to non-participation of many parameters in the training process. Consequently, analyzing the eigenvalue spectrum of Fisher matrices across multiple instances of \( \theta \) provides insights into the trainability and resilience of QNNs against barren plateaus. A neural network with higher trainability would exhibit reduced eigenvalue degeneracy.

The Fisher information matrix can be computed for the specific hyperparameters of the circuit. Following the approach outlined in Ref.~\cite{abbas2020power}, we generate a Gaussian-distributed dataset \( x \sim \mathcal{N}(\mu = 0, \sigma^2 = 1) \). The joint probability is then determined by evaluating the overlap between the prepared quantum state and the state produced by our quantum layer.

\begin{equation}
    P(y, x|\theta) = \braket{y|\psi (\theta, x)}, 
\end{equation}
where $y$ is the output state. By averaging over all $x$ and $y$, the Fisher information is calculated for any given $\theta$.

The metric's result consists of several images. Fig.~\ref{fig:QC}(a) shows distribution with several parameters having eigenvalues much greater than zero. This indicates that many of them are impactful, giving good trainability. In Fig.~\ref{fig:QC}(b), we can see a clear diagonal where every element is active and contributes to the circuit's performance. Lack of non-diagonal elements also shows that there is no pair of parameters that are entangled to the extent where we can replace them with a singular one.

As discussed in Ref.~\cite{larocca2021overparam}, certain quantum neural networks (QNNs) may exhibit reduced parameter efficiency due to over-parametrization. This effect is observed when, beyond a certain point, adding more parameters no longer increases the rank of the FIM, indicating that the circuit has reached a saturation point. At this stage, further parameter addition does not enhance expressivity and may lead to the risk of over-parametrization. Choosing the right number of parameters, enough to utilize the full capacity of the circuit but not too much, is the important issue that is resolved with the average FIM rank Fig.~\ref{fig:QC}(c).

\subsubsection{Expressivity analysis with Fourier series}\label{sec:appendix_fourier}

It has been demonstrated that the output of a parameterized quantum circuit can be represented as a truncated Fourier series. For a feature vector of length \( N \), this Fourier series is expressed as a function of the feature vector \( x \) and the trainable parameters $\theta$:

\begin{equation}
    f_\theta (x) = \sum\limits_{\omega_1 \in \Omega_1} ... \sum\limits_{\omega_N \in \Omega_N} c_{\omega_1, ... \omega_N}(\theta) e^{-i\omega x}
\end{equation}
where $\omega \in \left\{d_{-i}, ... 0, ..., d_i\right\}$.

In other words, the number of terms in the Fourier series is one plus twice the number of times the input is embedded in the circuit, denoted by \( d \). In this analysis, we assess the expressivity of the function \( f_{\theta}(x) \) by sampling over a uniform distribution of random values for each \( \theta_i \) within the interval \([0, 2\pi]\) and by selecting equidistant \( x \) values with a sampling frequency of \( d \).

This analysis centers on two primary considerations: the number of accessible terms in the Fourier series (the degree) and the accessibility of the coefficients for each term (the coefficients' expressivity). The expressivity is indicated by the number of non-zero coefficients in the Fourier spectrum—the higher the number, the greater the expressivity. 

Notably, this circuit demonstrates a high degree of expressivity, with $127$ non-zero coefficients out of $161$ possible terms (around $79\%$) as shown in Fig.~\ref{fig:QC}(d).

\subsection{Training curves}

\begin{figure*}[!htbp]
\centering
 \includegraphics[width=.9\linewidth]{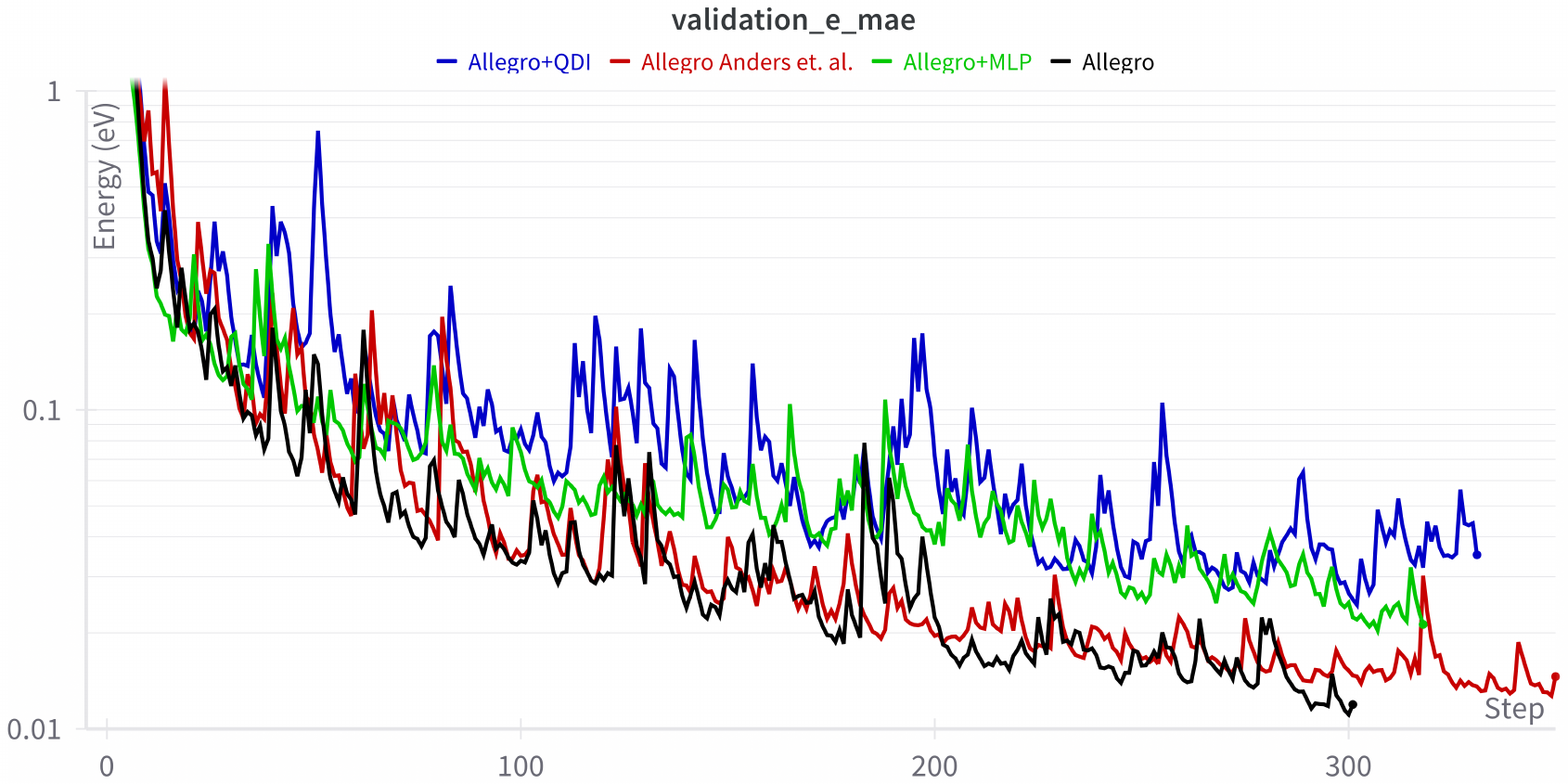}
 \caption{Training curves of all Allegro variants on QM9 dataset using the best hyperparameters, as achieved from our hyperparameter optimization. Among the variants, we include the Allegro Anders et al. training curve, which corresponds to QM9 Allegro hyperparameters as reported in \cite{allegro_paper}. To smooth the resulting curves we use the time-weighted exponential moving average (EMA) \cite{time_EMA} smoothing technique with 0.75 smoothing factor.} 
 \label{fig:qm9_curves}
\end{figure*}

\begin{figure*}[h!]
    \centering
    \begin{subfigure}[b]{0.48\textwidth}  
        \centering
        \includegraphics[width=\textwidth]{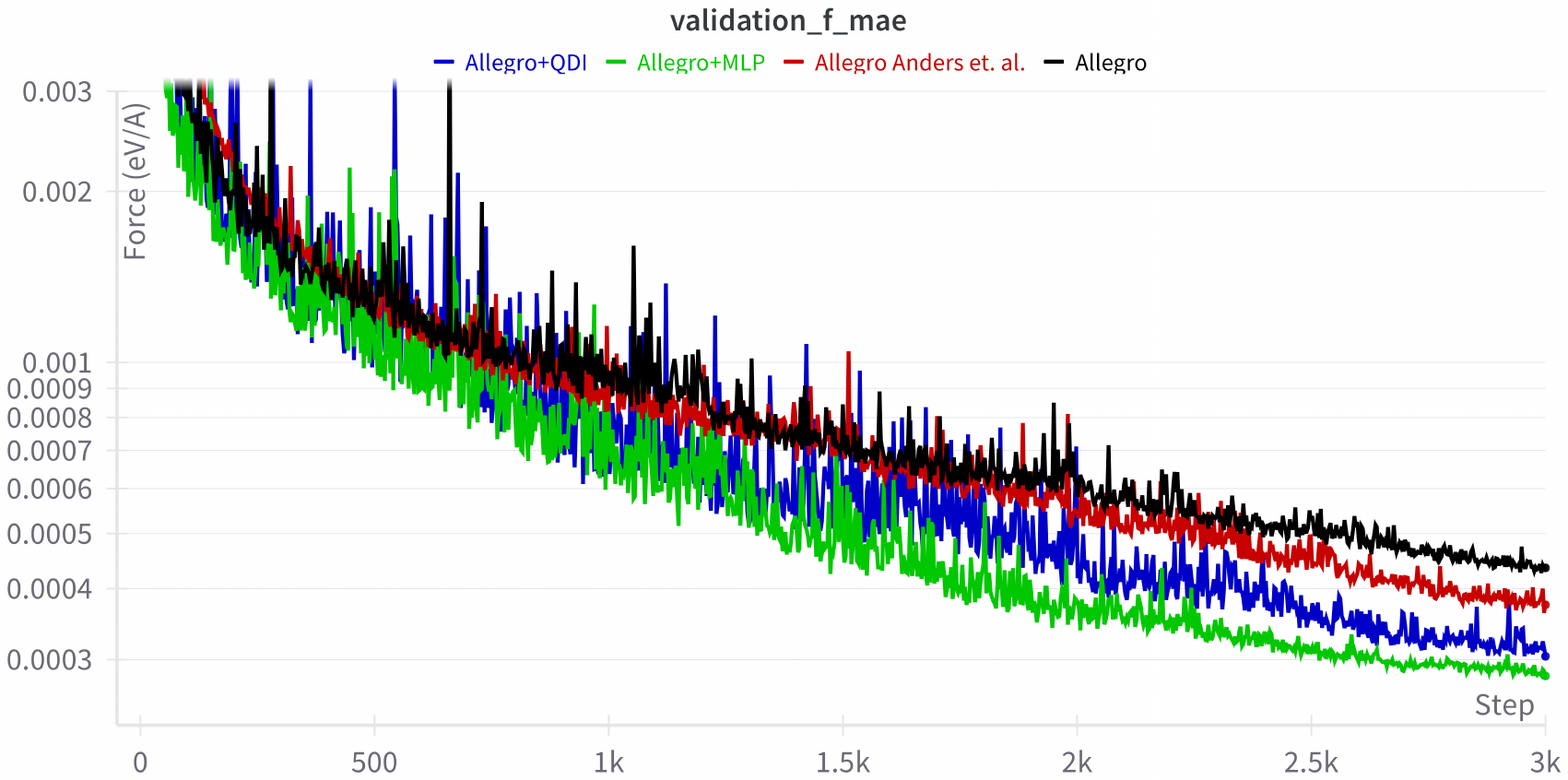}
        \caption{rMD17-benzene validation F MAE}
        \label{fig:fig1}
    \end{subfigure}
    \hfill
    \begin{subfigure}[b]{0.48\textwidth}
        \centering
        \includegraphics[width=\textwidth]{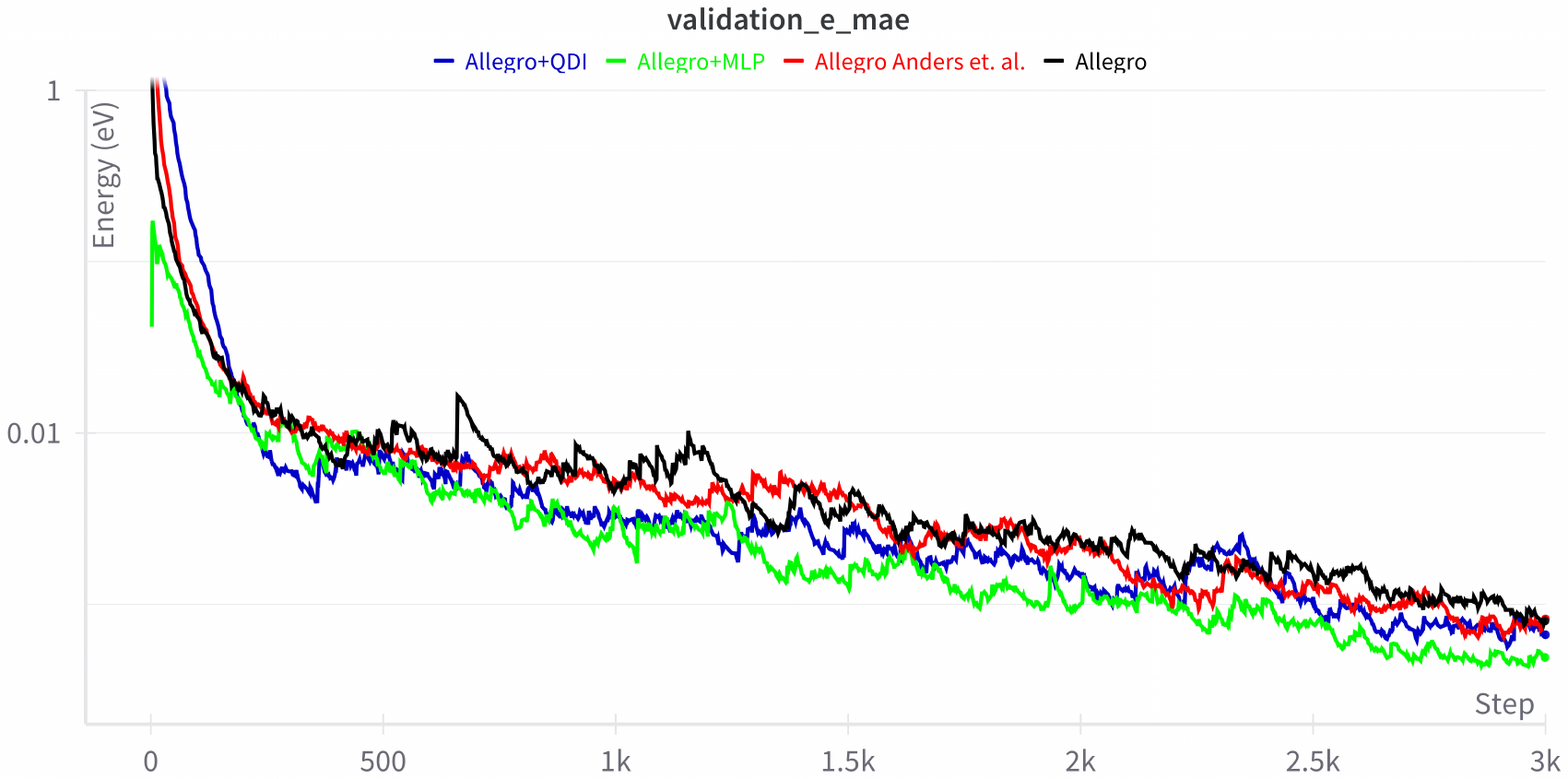}
        \caption{rMD17-benzene validation E MAE}
        \label{fig:fig2}
    \end{subfigure}

    
    \begin{subfigure}[b]{0.48\textwidth}
        \centering
        \includegraphics[width=\textwidth]{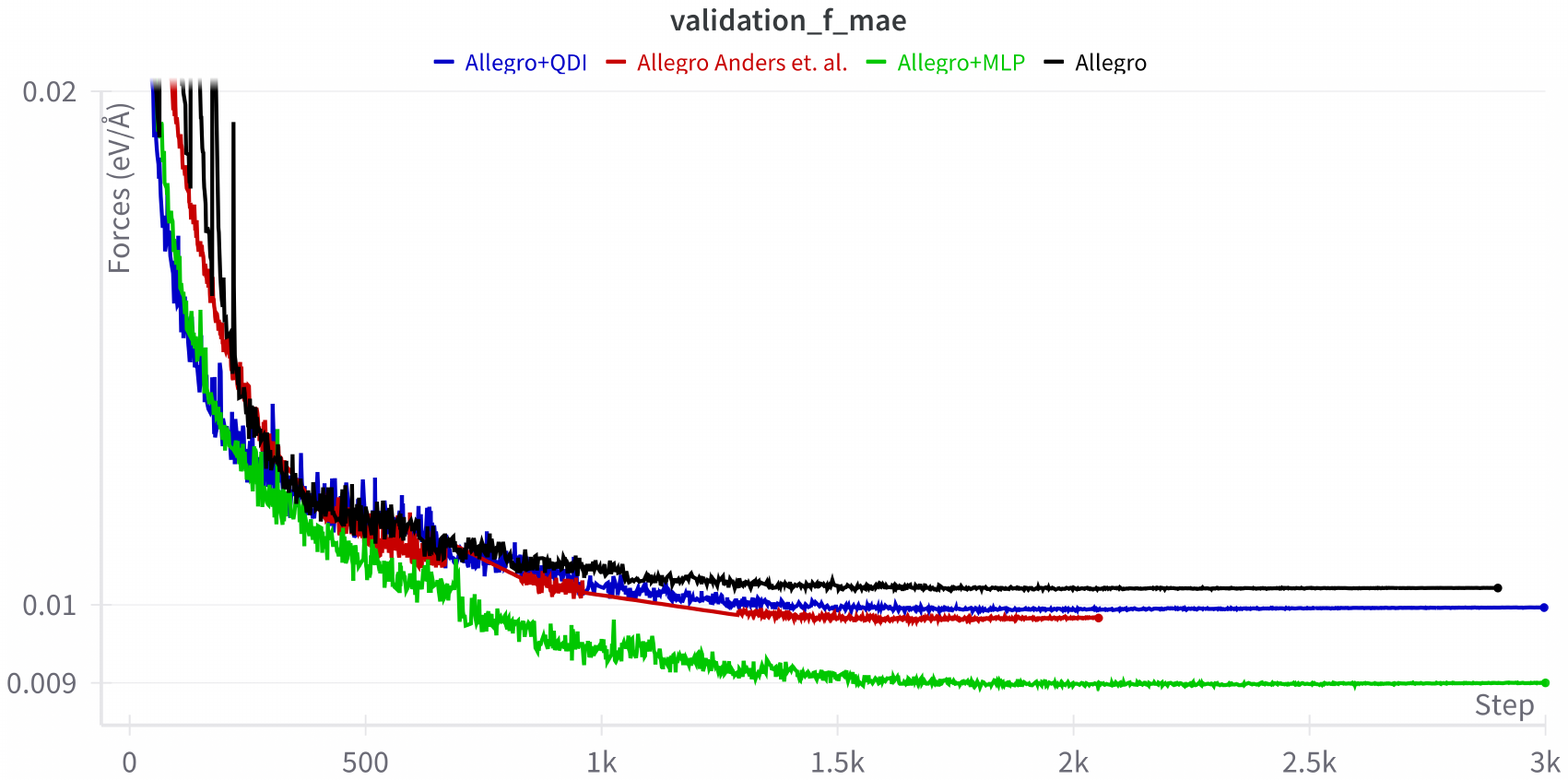}
        \caption{rMD17-aspirin validation F MAE}
        \label{fig:fig3}
    \end{subfigure}
    \hfill
    \begin{subfigure}[b]{0.48\textwidth}
        \centering
        \includegraphics[width=\textwidth]{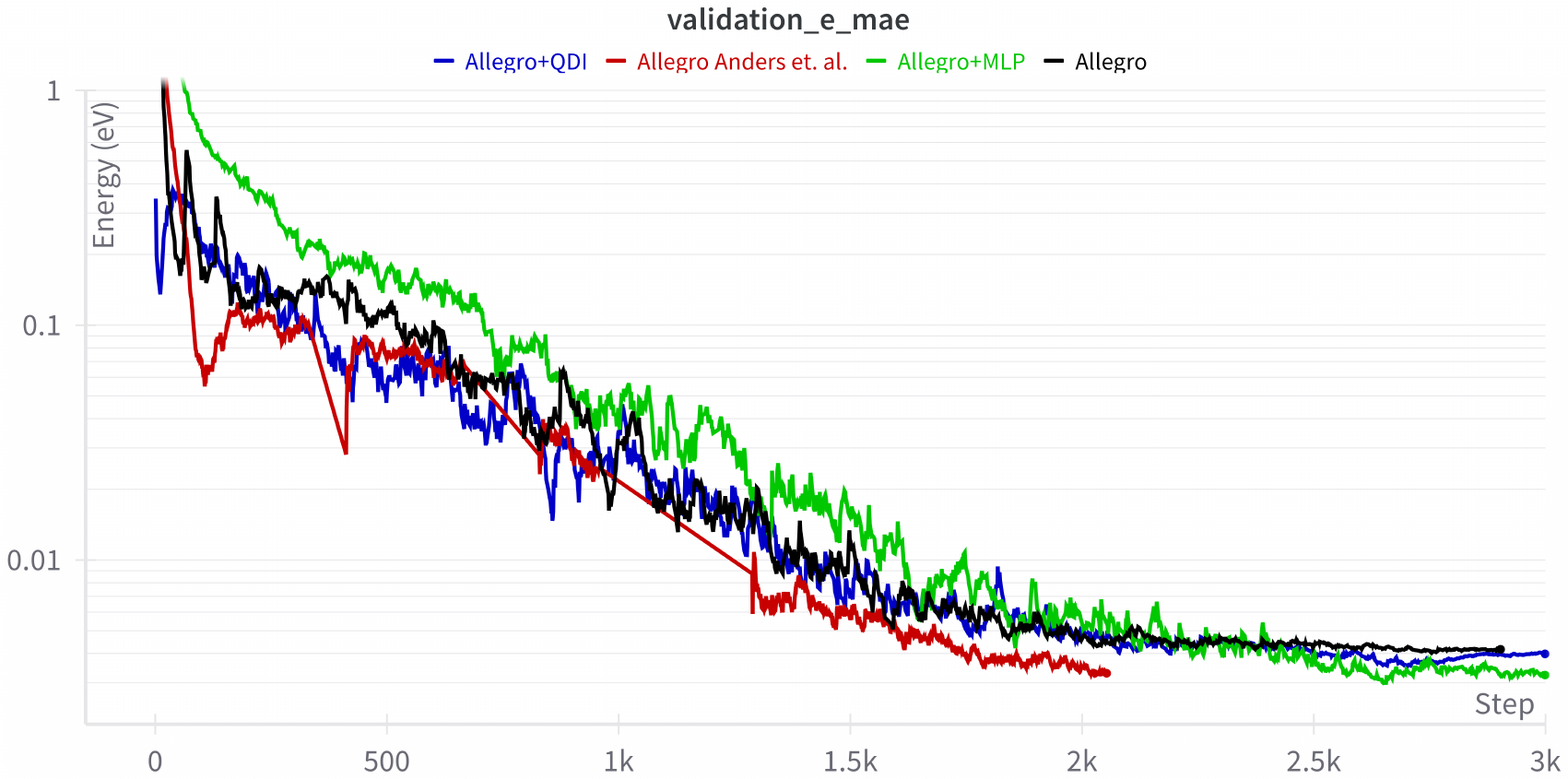}
        \caption{rMD17-aspirin validation E MAE}
        \label{fig:fig4}
    \end{subfigure}

    
    \begin{subfigure}[b]{0.48\textwidth}
        \centering
        \includegraphics[width=\textwidth]{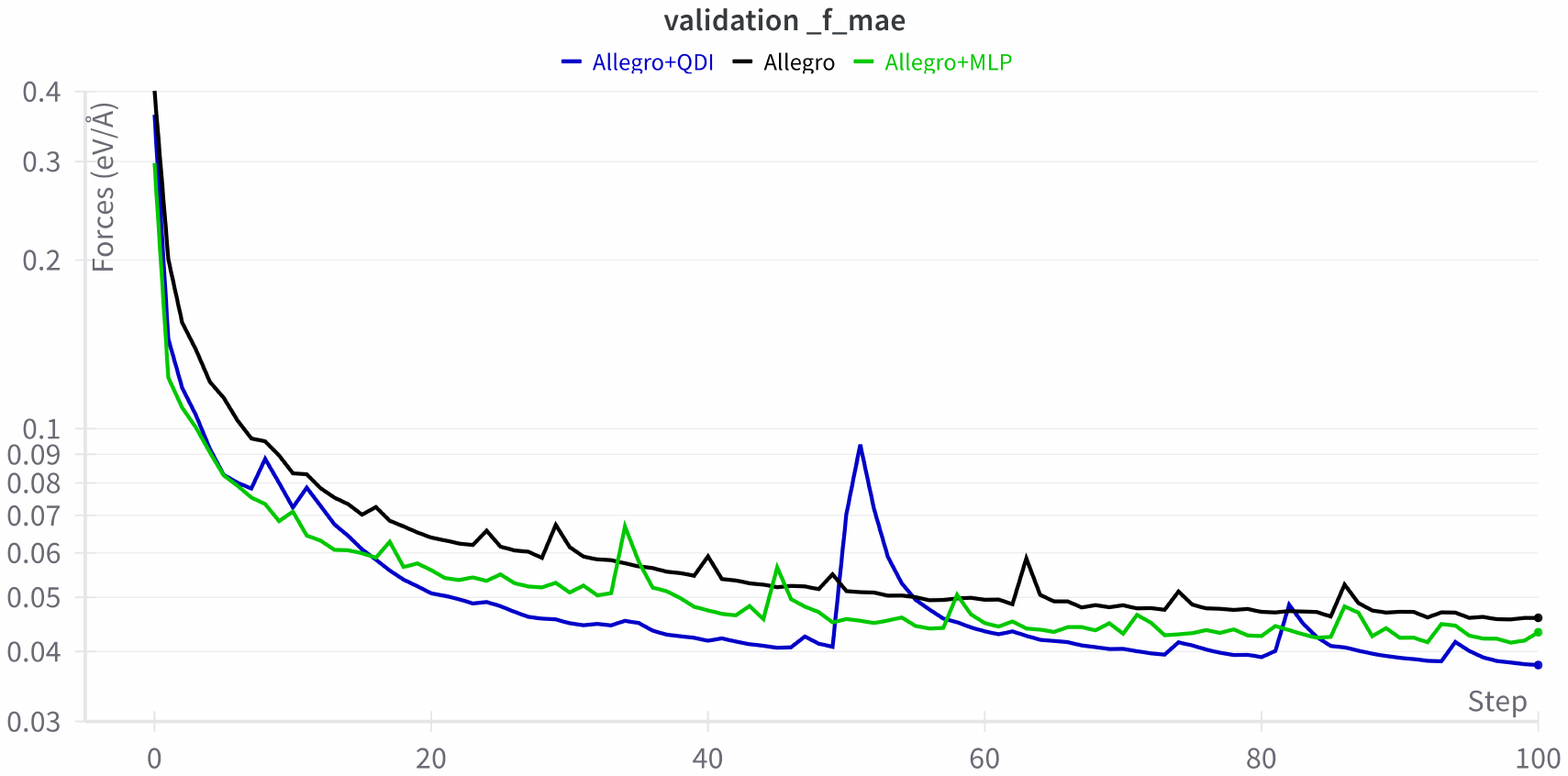}
        \caption{CuLi validation F MAE}
        \label{fig:fig5}
    \end{subfigure}
    \hfill
    \begin{subfigure}[b]{0.48\textwidth}
        \centering
        \includegraphics[width=\textwidth]{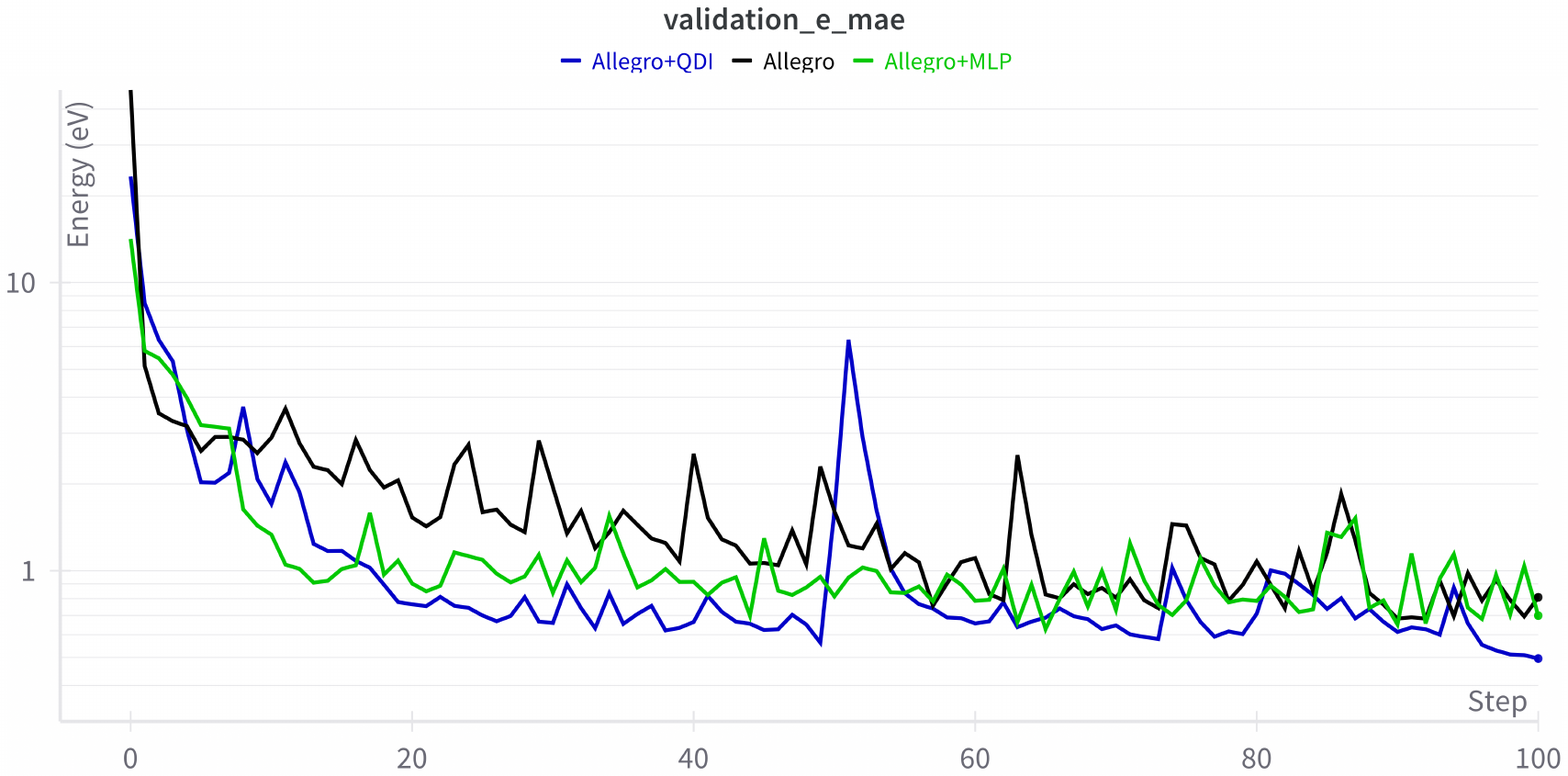}
        \caption{CuLi validation E MAE}
        \label{fig:fig6}
    \end{subfigure}
    \caption{Training curves for all Allegro variants including Allegro with the parameters reported by Anders et al. \cite{allegro_paper}. We report mean absolute errors both forces and energies for all datasets up to 3000 epochs (except from QM9 and CuLi which is up to 300 and 100 epochs respectively). Moreover, for the energy sub-figures (b) and (d) we use the time weighted EMA \cite{time_EMA} smoothing technique with smoother factors $0.9$ and $0.8$ respectively. All graphs are plotted with semi-logarithmic scaling in y-axis.}
    \label{fig:3x2grid}
\end{figure*}

\begin{figure*}[!htbp]
\centering
 \includegraphics[width=.9\linewidth]{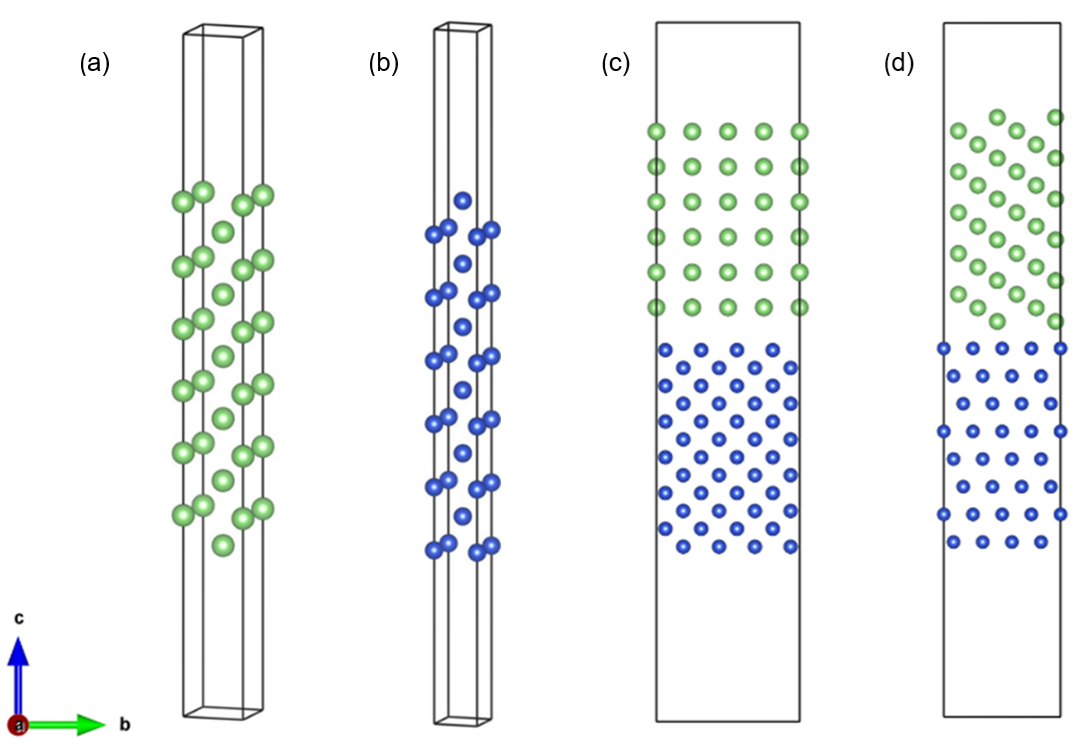}
 \caption{ An example generated (a) Li (110) and (b) Cu (110) slabs with a thickness of 20 \AA{} and 30 \AA{}  and vacuum spacing of 15 \AA{}. Interface generated between the (c) Li (110) and Cu (110) slabs and (d) Li (111) and Cu (111) slabs with the same vacuum spacing of 15 \AA{}.  } 
 \label{fig:CuLi_systems}
\end{figure*}

\end{document}